\documentclass[12pt]{article}
\usepackage{latexsym}
\begin{document}
\vskip .5cm
\begin{center}
{\large \bf SYMMETRIES AND MASS PREDICTIONS OF THE  SUPERSYMMETRIC QUARK MODEL\\}

\vskip 1.1cm
\large {\bf Sultan  \ Catto$^{\dag}$}\\
\vskip .2in
\normalsize Physics Department\\The Graduate School and University Center \\
The City University of New York\\
365 Fifth Avenue\\
New York, NY 10016\\ and\\
Center for Theoretical Physics\\ The Rockefeller University\\
1230 York Avenue\\ New York, NY 10021-6399\\ 

\end{center}
\vskip .1cm
\begin{abstract}
QCD justification of $SU(m/n)$ supergroups are shown to provide a basis for the existence of an approximate hadronic supersymmetry. Effective Hamiltonian of the relativistic quark model is derived, leading to hadronic mass formulae in remarkable agreement with experiments. Bilocal approximation to hadronic structure and incorporation of color through octonion algebra (based on quark-antidiquark symmetry) is also shown to predict exotic diquark-antidiquark ($D-\bar{D})$ meson states. A minimal supersymmetric scheme based on $SU(3)^c \times SU(6/1)$ that excludes exotics is constructed. Symmetries of three quark systems and possible relativistic formulation of the quark model through the spin realization of Wess Zumino algebra is presented.
\end{abstract}
\vskip 1cm
\begin{flushleft}
$^\dag$ Work supported in part by DOE contract No. DE-AC-0276 ER 03074 and 03075, and PSC-CUNY
Research Awards.
\end{flushleft}
\newpage

\newpage
\begin{center}

{\bf Introduction}

\end{center}

$\;\;\;$In $1966$ Miyazawa, in a series of papers$^{\cite{m1}}$, extended the $SU(6)$ group 
to the supergroup $SU(6/21)$ that could be generated by constituent quarks and 
diquarks that could be transformed to each other. In particular, he found the
following 
\begin{itemize}
\item a general definition
of $SU(m/n)$ superalgebras, expressing the symmetry between $m$ bosons and $n$ fermions,
with Grassman-valued parameters.
\item  a derivation of the super-Jacobi identity.
\item the relation of the baryon mass
splitting to the meson mass splitting through the new mass formulae. 
\end{itemize}
This work contained the first classification of superalgebras (later rediscovered 
by mathematicians in the seventies). Because of the field-theoretic prejudice 
against $SU(6)$, Miyazawa's work was generally
ignored. Supersymmetry was, of course, rediscovered in the 
seventies within the dual resonance model by Ramond$^{\cite{ram}}$, Neveu and Schwarz$^{\cite{nev}}$
$(1971)$. Golfand and Likhtman$^{ \cite{gol}}$, and independently
Akulov and Volkov$^{\cite{vol}}$, proposed
the extension of the Poincar\'e group to the super-Poincar\'e group. Examples of
supersymmetric
field theories were given and the general method based on the super-Poincar\'e group was discovered
by Wess and Zumino$^{ \cite{wess}}$ $(1974)$. The super-Poincar\'e group allowed transformations between
fields associated
with different spins $0$, $\frac{1}{2}$ and $1$. The Coleman-Mandula theorem was amended in
1975 by Haag,
Lopuszanski and Sohnius to allow ${\bf super-Poincar\grave e ~ group \times G_{int}}$ as the
maximum symmetry 
of the
S-matrix. Unfortunately, $SU(6)$ symmetry was still forbidden.

\begin{center}

{\bf  SU(6) and Hadronic Supersymmetry}

\end{center}

How do we interpret the symmetries of the QCD spectrum in this light?  In the ultraviolet, the
running
coupling constant tends to zero and quarks behave like free point particles. Thus an
approximate conformal symmetry exists, allowing spin to be
conserved separately from orbital angular momentum. Thus spin behaves as an internal
quantum number; this makes an $SU(6)$ symmetry possible, since the quarks
are almost free Dirac particles. Single
vector-gluon exchange breaks this symmetry; thus, as shown by 
Glashow, Georgi and deRujula$^{ \cite{geo}}$, the mass-degeneracy of hadrons of  
different spins is lifted by a hyperfine-interaction term.

Here is the main point. In the infrared we expect 
confinement to set in.  The quark-antiquark potential becomes
proportional to the distance. Careful studies of quarkonium spectra and lattice-gauge 
calculations show that at large separation the quark
forces become spin-independent.  QCD is also flavor independent. We therefore 
find approximate spin- and flavor-independent quark binding forces; these are completely
consistent with $SU(6)$ symmetry. This is not an exact symmetry, but is a good starting point, before spin and
flavor effects are included.

There is good phenomenological evidence that in a rotationally
excited baryon a quark-diquark $(q-D)$ structure is favored over a three-quark $(qqq)$ 
structure$^{ \cite{l1}, \cite{anse},\cite{cg1}}$.  Eguchi$^{\cite{egu}}$ had shown that it is energetically favorable for the three quarks in a baryon to form a linear structure with a quark on one end and bilocal structure $qq$ at the other end. Similarly,  Johnson and
Thorn$^{\cite{jo}}$ had shown that the string-like hadrons may be pictured as vortices of color flux lines which terminate on concentration of color at the end points. A baryon with three valence quarks would be arranged as a linear chain of molecule where the largest angular momentum for a state of a given mass is expected when two quarks are at one end, and the third is at the other: At large spin, two of the quarks form
a diquark at one end of the string, the remaining quark being at the other. Regge trajectories for mesons and baryons are closely parallel; both have a
slope of about $0.9(GeV)^{-2}$.   If the quarks are light, the underlying quark-diquark symmetry leads to a
Miyazawa symmetry between mesons and baryons. Thus we studied QCD
with a weakly broken supergroup
$SU(6/21)$. Note that the fundamental
theory is not supersymmetric. For quarks, the generators of the
Poincar\'e group and those of the color group $SU(3)^c$ commute. It is only the
effective Hamiltonian 
which exhibits an approximate
supersymmetry among the bound states $q\bar{q}$ and $qD$. 

Under the color group $SU^C(3)$, meson $q \bar{q}$  and diquark ($D=qq$) states
transform as$^{\cite{cg1},\cite{cg2}}$

\begin{equation}
qq:~~{\bf{3\times 3=\bar{3}+ 6}}~~;~~~~q\bar{q}:~~{\bf{3\times \bar{3}=1+8}}
\end{equation}
and under the spin flavor $SU(6)$ they transform as

\begin{equation}
qq:~~{\bf{6\times 6=15+21}}~~;~~~~q\bar{q}:~~{\bf{6\times \bar{6}=1+35}}
\end{equation}

Dimensions of internal degrees of quarks and diquarks are shown in the following table:

\begin{center}
\begin{tabular}{|l|c|c|c|}\hline
 & $SU_f(3)$ & $SU_s(2)$ & $dim.$  \\ \hline \hline
$q$  &  $ \Box $  & $s=1/2$ & $3\times2=6$ \\ \hline
$D$ & 
$  
\begin{array}{c}
 ~~~~\Box~~\Box \\
  \\
 \Box \\
\Box
  \end{array} 
\nonumber
$ 
& 
$  
\begin{array}{c}
 s=1 \\
  \\
 s=0 \\
\\
  \end{array} 
\nonumber
$ 

 & 
$  
\begin{array}{c}
 6\times3=18 \\
  \\
 3\times1=3 \\
\\
  \end{array} 
\nonumber
$ 

\\\hline

\end{tabular}
\end{center}

If one writes $qqq$ as $qD$, then the quantum numbers of $D$ are ${\bf \bar{3}}$ for color since when combined
with $q$ must give a color singlet, and ${\bf{21}}$ for spin-flavor since combined with color must give antisymmetric wavefunctions. The quantum numbers for $\bar{q}$ are for color, ${\bf \bar{3}}$, and for spin-flavor, ${\bf \bar{6}}$. Thus $\bar{q}$ and $D$ have the same quantum numbers (color forces can not distinguish between $\bar{q}$ and $D$). Therefore there is a dynamic supersymmetry in hadrons with supersymmetric partners

\begin{equation}
\psi=\left(
\begin{array}{c}
\bar{q} \\
D
\end{array}
\right)~,~~~~~\bar{\psi}= ( q~~~\bar{D})
\end{equation}

We can obtain all hadrons by combining $\psi$ and $\bar{\psi}$: mesons are $q\bar{q}$, baryons are $qD$, and exotics are $D\bar{D}$ states. Inside rotationally excited baryons, QCD leads to the formation of diquarks well separated from the remaining quark. At this separation the scalar, spin-independent, confining part of the effective QCD potential is dominant. Since QCD forces are also flavor-independent, the force between the quark $q$ and the diquark $D$ inside an excited baryon is essentially the same as the one between $q$ and the antiquark $\bar{q}$ inside an excited meson. Thus the approximate spin-flavor independence of hadronic physics expressed by $SU(6)$ symmetry is extended to $SU(6/21)$ supersymmetry through a symmetry between $\bar{q}$ and $D$, resulting in parallelism of mesonic and baryonic Regge trajectories.   

\newpage
\begin{center}
{\bf QCD Justification of $U(6/21)$ Supersymmetry and Its Breaking}
\end{center}

The approximate group $SU(6)$ and its supersymmetric extension to $SU(6/21)$ was justified within the standard theory of colored quarks interacting through gluons that are associated with a color group $SU(3)^c$. The justification of $SU(6)$ and the derivation of its breaking was given by Georgi, Glashow and de Rujula$^{\cite{geo}}$. We extended their approach through use of supergroups of type $SU(m/n)$ which provided a basis for the existence of an approximate hadronic supersymmetry$^{\cite{cg1},\cite{cg2}}$. 

We shall first discuss the validity domain of $SU(6/21)$ supersymmetry. The diquark structure with spin $s=0$ and $s=1$ emerges in inelastic inclusive lepton-baryon collisions with high impact parameters that excite the baryon rotationally, resulting in inelastic structure functions based on point-like quarks and diquarks instead of three point-like quarks, In this case both mesons and baryons are bilocal with large separation of constituents.

Also, there is a symmetry between color antitriplet diquarks with $s=0$ and $s=1$ and color antitriplet antiquarks with $s=\frac{1}{2}$. This is only possible if the force between quark $q$ and antiquark $\bar{q}$, and also between $q$ and diquark $D$ is mediated by a zero spin object that sees no difference between the spins of $\bar{q}$ and $D$. The object can be in color states that are either singlet or octet since $q$ and $D$ are both triplets. Such an object is provided by scalar flux tubes of gluons that dominate over the one gluon exchange at large distances. Various strong coupling approximations to QCD, like lattice gauge theory$^{\cite{d},\cite{e}}$,  't Hooft's $\frac{1}{N}$ approximation$^{\cite{f}}$ when $N$, the number of colors, is very large, or the elongated bag model$^{\cite{jo}}$ all give a linear potential between widely separated quarks and an effective string that approximates the gluon flux tube. In such a theory it is energetically favorable for the three quarks in a baryon to form a linear structure with a quark in the middle and two at the ends, or, for high rotational excitation, a bilocal linear structure (diquark) at one end and a quark at the other end. In order to illustrate these points we start with the suggestion of Johnson and Thorn$^{\cite{jo}}$ that the string-like hadrons may be pictured as vortices of color flux lines which terminate on concentration of color at the end points. The color flux connecting opposite ends is the same for mesons and baryons giving an explanation for the same slope of meson and baryon trajectories$^{\cite{cg1}}$.

To construct a solution which yields a maximal angular momentum for a fixed mass we consider a bag with elongated shape rotating about the center of mass with an angular frequency $\omega$. Its ends have the maximal velocity allowed, which is the speed of light $(c=1)$. Thus, a given point inside the bag, at a distance $r$ from the axis of rotation moves with a velocity

\begin{equation}
v=\vec{\omega} \cdot \vec{r} = \frac{2r}{L}     \label{eq:A}
\end{equation}
where $L$ is the length of the string. In this picture the bag surface will be fixed by balancing the gluon field pressure against the confining vacuum pressure $B$, which (in analogy to electrodynamics) can be written in the form

\begin{equation}
\frac{1}{2} \sum_{\alpha=1}^{8} (E_\alpha^2 -B_\alpha^2) = B        \label{eq:B}
\end{equation}

Using Gauss' law the color electric field $E$ through the flux tube connecting the color charges at the ends of the string is given by

\begin{equation}
\int {\vec{E}}_\alpha \cdot d\vec{S} = E_\alpha A = g\frac{1}{2} \lambda_\alpha        \label{eq:C}
\end{equation}
where$A(r)$ is the cross-section of the flux tube at distance $r$ from the center and $g\frac{1}{2}$ is the color electric charge which is the source of $E_\alpha$. By analogy with classical electrodynamics the color magnetic field ${\vec{B}}_\alpha(r)$ associated with the rotation of the color electric field is

\begin{equation}
{\vec{B}}_\alpha(r)= \vec{v}(r) \times {\vec{E}}_\alpha(r)        \label{eq:D}
\end{equation}
at a point moving with a velocity $\vec{v}(r)$. For the absolute values this implies

\begin{equation}
B_\alpha=v~E_\alpha               \label{eq:E}
\end{equation}
because $\vec{v}(r)$ is perpendicular to ${\vec{E}}_\alpha(r)$. Using last three equations together with

\begin{equation}
<\sum_{\alpha=1}^{8} (\frac{1}{2} \lambda_\alpha)^2>=\frac{4}{3}
\end{equation}
for the $SU(3)^c$ triplet in Eq.(\ref{eq:B}) we obtain for the crosssection of the bag

\begin{equation}
A(r)= \sqrt{\frac{2}{3B}} ~g~\sqrt{1-v^2}              \label{eq:F}
\end{equation}
which shows the expected Lorentz contraction.

The total energy $E$ of the bag

\begin{equation}
E= E_q +E_G +BV
\end{equation}
is the sum of the quark energy $E_q$, the gluon field energy $E_G$ and the volume energy of the bag, $BV$. Because the quarks at the ends move with the a speed close to speed of light, their energy is simply given by

\begin{equation}
E_q=2p
\end{equation}
where $p$ is the momentum of a quark, a diquark or an antiquark, respectively. For the gluon energy, by analogy with electrodynamics, one obtains from Eqs.(\ref{eq:C}-\ref{eq:E}) the result
  
\begin{eqnarray}
E_G=\frac{1}{2} \int d^3x \sum_{\alpha=1}^{8} (E_\alpha^2 + B_\alpha^2) ~~~~~~~~~~~~~~~~~~~~~~~~~~\nonumber \\
= \sqrt{\frac{2}{3}} g \sqrt{B} L \int_0^1 dv \frac{1+v^2}{\sqrt{1-v^2}} = \sqrt{\frac{2}{3}}g\sqrt{B} L\frac{3\pi}{4}      \label{eq:G}
\end{eqnarray}
and for the volume energy

\begin{eqnarray}
BV= 2B\int_0^{\frac{L}{2}} A(r)~dr=2B\int_0^1 \sqrt{\frac{2}{3B}} g \sqrt{1-v^2} \frac{L}{2} dv \nonumber \\
=\sqrt{\frac{2}{3}} g \sqrt{B} L \frac{\pi}{4} = \frac{BA(0)L\pi}{4}~~~~~~~~~~~~~~~~~~~~~~~      \label{eq:H}
\end{eqnarray}
 
It is obvious from Eq.(\ref{eq:G}) that the gluon field energy is proportional to the length $L$ of the bag. The gluon field energy and the volume energy of the bag together correspond to a linear rising potential of the form

\begin{equation}
V(L)=E_G +BV =bL      \label{eq:I}
\end{equation}
where 

\begin{equation}
b=\sqrt{\frac{2B}{3}} g \pi
\end{equation}
The total angular momentum $J$ of this classical bag is the sum of the angular momenta of the quarks at the two ends

\begin{equation}
J_q=pL             \label{eq:J}
\end{equation}
and the angular momentum $J_G$ of the gluon field. From Eq.(\ref{eq:D}) we get 

\begin{equation}
{\vec{E}}_\alpha \times {\vec{B}}_\alpha= \vec{v} E_\alpha^2 
\end{equation}
for the momentum of the gluon field, and hence
  
\begin{eqnarray}
J_G=| \int_{bag} d^3\vec{r} \sum_{\alpha=1}^{8} \vec{r} \times ({\vec{E}}_\alpha \times{\vec{B}}_\alpha)| = 2\int_0^{\frac{L}{2}} dr A(r) r v E_\alpha^2 \nonumber \\
= \frac{16}{3L} g^2 \int_0^{\frac{L}{2}} \frac{r^2 dr}{A(r)}=\sqrt{\frac{2}{3}} g \sqrt{B} L^2 \frac{\pi}{4}~~~~~~~~~~~~~~~~~~
\end{eqnarray}
where we have used Eq.(\ref{eq:A}) and Eq.(\ref{eq:C}) in the third step. We can now express the total energy of the bag in terms of angular momenta. Putting these results back into expressions for $E_q$ and $E_G$, we arrive at

\begin{equation}
E_q = \frac{2J_q}{L},~~~~~~~~~~E_G= \frac{3J_G}{L}
\end{equation}
so that the bag energy now becomes

\begin{eqnarray}
E= \frac{2J_q}{L}+ \frac{3J_G}{L}+ \sqrt{\frac{2B}{3}} L g \frac{\pi}{L}=\frac{2J_q+4J_G}{L} \nonumber \\
=\frac{2(J+J_G)}{L}= \frac{1}{L} (2J+\sqrt{\frac{2}{3}}g\sqrt{B} L^2 \frac{\pi}{2})      \label{eq:K2}
\end{eqnarray}
Minimizing the total energy for a fixed angular momentum with respect to the length of the bag, $\frac{dE}{dt}=0$ gives the relation

\begin{equation}
- \frac{2J}{L^2}+ \sqrt{\frac{2}{3}} g \sqrt{B} \frac{\pi}{2} =0
\end{equation}
so that

\begin{equation}
L^2= \frac{4J}{g\pi} \sqrt{\frac{3}{2B}}
\end{equation}
Re-inserting this into Eq.(\ref{eq:K2}) we arrive at

\begin{equation}
E =2 \sqrt{Jg\pi} (\frac{2B}{3})^{\frac{1}{4}}
\end{equation}
or

\begin{equation}
J= (\sqrt{\frac{3}{2B}} \frac{1}{4\pi g})E^2=(\sqrt{\frac{3}{2B}} \frac{1}{8\pi^{\frac{3}{2}}}\frac{1}{\sqrt{\alpha_s}})E^2=\alpha^{'}(0) M^2  
\end{equation}
where $M=E$ and we used $\alpha_s=\frac{g^2}{4\pi}$, the unrationalized color gluon coupling constant. We can now let $\alpha^{'}(0)$ defined by the last equation which is the slope of the Regge trajectory as

\begin{equation}
\alpha^{'}(0)=\sqrt{\frac{3}{2B}} \frac{1}{8\pi^{\frac{3}{2}}}\frac{1}{\sqrt{\alpha_s}} =\frac{1}{4b}   \label{eq:M}
\end{equation}
where $b$ was defined in Eq.({\ref{eq:I}). 

The parameters $B$ and $\alpha_s$ have been determined$^{\cite{chodos}.\cite{chodos2}}$ using the experimental information from the low lying hadron states: $B^{\frac{1}{4}}=0.146~GeV$ and $\alpha_s=0.55~GeV$.  If we use these values in Eq.(\ref{eq:M}) we find

\begin{equation}
\alpha^{'}(0)=0.88~(GeV)^{-2}
\end{equation}
in remarkable agreement with the slope determined from experimental data which is about $0.9~(GeV)^{-2}$.

The total phenomenological non-relativistic potential then is the well known superposition of the Coulomb-like and confining potentials $V(r)= \frac{a}{r} + br$ where $r=|\vec{r}_1-\vec{r}_2|$ is the distance between $q$ and $\bar{q}$ in a meson, or betwee $q$ and $D$ in a baryon with high angular momentum. This is verified in lattice QCD to a high degree of accuracy$^{\cite{otto}}$  ($a=\frac{-c\alpha_c}{r}$, where $c$ is the color factor and $\alpha_c$ the strong coupling strength).

It is interesting to know that all this is related very closely to the dual strings. Indeed we can show that the slope given in Eq.(\ref{eq:M}) is equivalent to the dual string model formula for the slope if we associate the "proper tension" in the string with the proper energy per unit length of the color flux tube and the volume. By proper energy per unit length we mean the energy per unit length at a point in the bag evaluated in the rest system of that point. This will be

\begin{equation}
T_0= \frac{1}{2} \sum_\alpha E_\alpha^2 A_0 + BA_0
\end{equation}
The fact that $ \frac{1}{2} \sum_\alpha E_\alpha^2 =B$ in the rest sytem gives
 
\begin{equation}
T_0=2BA_0
\end{equation}
where $A_0$ is the cross-sectional area of the bag. If in Eq.(\ref{eq:F}) we let $A=A_0$ when $v=0$, then using

\begin{equation}
A_0=\sqrt{\frac{2}{3B}} g
\end{equation}
we find

\begin{equation}
T_0= 2\sqrt{\frac{2}{3} } g \sqrt{B} = 4 \sqrt{\frac{2\pi}{3}} \sqrt{\alpha_s}\sqrt{B}    \label{eq:N}
\end{equation}
for the proper tension. In the dual string the slope and proper tension are related by the formula$^{\cite{goddard}}$ 

\begin{equation}
T_0=\frac{1}{2\pi \alpha^{'}}   \label{eq:P}
\end{equation}
so the slope is

\begin{equation}
\alpha^{'}= \frac{1}{8} \sqrt{\frac{3}{2}} \frac{1}{\pi^{\frac{3}{2}} } \frac{1}{\sqrt{\alpha_s}} \frac{1}{\sqrt{B}}
\end{equation}
which is identical to the earlier formula we produced in Eq.(\ref{eq:M}). 

It would appear from Eq.(\ref{eq:N}) that the ratio of volume to field energy would be one-to-one in one space dimension in contrast to the result one-to-three which holds for a three dimensional bag$^{\cite{chodos3}}$.  However, the ratio one-to-one is true only in the rest system at a point in the bag, and each position along the $x$-axis is of course moving with a different velocity. Indeed we see from Eq.(\ref{eq:G}) and Eq.(\ref{eq:H}) that the ratio of the total volume energy to the total field energy is given by one-to three in conformity with the virial theorem$^{\cite{chodos3}}$. 

In the string model of hadrons we have $E^2\sim J $ between the energy and the angular momentum of the rotating string. If we denote by $\rho(r)$ the mass density of the string, and by $v$ and $\omega$ its linear and angular velocities, respectively, the energy and the angular momentum of the rotating string are given by

\begin{equation}
E= 2\int \frac{\rho(r)}{\sqrt{1-\omega^2 r^2}}  dr =\frac{2}{\omega} \int_0^1 \frac{\rho(v)}{\sqrt{1-v^2}}  dv
\end{equation}
and

\begin{equation}
J= 2\int \frac{\rho(r)}{\sqrt{1-\omega^2 r^2}} r^2 \omega dr =\frac{2}{\omega^2} \int_0^1 \frac{\rho(v)}{\sqrt{1-v^2}} v^2 dv
\end{equation}
and hence the relation 

\begin{equation}
E^2 \propto J.  \label{eq:Q}
\end{equation}

If the string is loaded with mass points at its ends, they no longer move with speed of light, however, the above relation still holds approximately for the total energy and angular momentum of the loaded string. 

We now look at various ways of partitioning of the total angular momentum into two subsystems. Figures (a), (b) and (c) show the possible configurations of three quarks in a baryon. 

\begin{center}

\unitlength=.3mm
\begin{picture}(240,180)(0,0)
\put(70,20){\line(0,3){120}}
\put(70,20){\line(20,0){40}}
\put(70,20){\line(-20,0){40}}
\put(30,20){\circle*{3.7}}
\put(110,20){\circle*{3.7}}
\put(70,140){\circle*{3.7}}
\put(20,20){\makebox(0,0){$q_1$}}
\put(120,20){\makebox(0,0){$q_2$}}
\put(70,150){\makebox(0,0){$q_3$}}
\put(70,10){\makebox(0,0){$J=0$}}
\put(70,-10){\makebox(0,0){$\bf {Fig.(a)}$}}
\put(110,80){\makebox(0,0){$J=J_1+J_2$}}
\end{picture}
\end{center}

\begin{center}

\unitlength=.3mm
\begin{picture}(240,180)(0,0)
\put(70,20){\line(0,3){60}}
\put(70,20){\line(20,0){60}}
\put(70,20){\line(-20,0){60}}
\put(10,20){\circle*{3.7}}
\put(130,20){\circle*{3.7}}
\put(70,80){\circle*{3.7}}
\put(0,20){\makebox(0,0){$q_1$}}
\put(140,20){\makebox(0,0){$q_2$}}
\put(70,90){\makebox(0,0){$q_3$}}
\put(70,10){\makebox(0,0){$J=J_1$}}
\put(70,-10){\makebox(0,0){$\bf {Fig.(b)}$}}
\put(100,50){\makebox(0,0){$J=J_2$}}
\end{picture}
\end{center}

\begin{center}

\unitlength=.3mm
\begin{picture}(240,180)(0,0)
\put(70,20){\line(0,3){20}}
\put(70,20){\line(20,0){90}}
\put(70,20){\line(-20,0){90}}
\put(-20,20){\circle*{3.7}}
\put(160,20){\circle*{3.7}}
\put(70,40){\circle*{3.7}}
\put(-30,20){\makebox(0,0){$q_1$}}
\put(170,20){\makebox(0,0){$q_2$}}
\put(70,50){\makebox(0,0){$q_3$}}
\put(70,10){\makebox(0,0){$J=J_1+J_2$}}
\put(70,-10){\makebox(0,0){$\bf {Fig.(c)}$}}
\put(90,30){\makebox(0,0){$J=0$}}
\end{picture}
\end{center}
\vskip 2cm
If we put the proportionality constant in Eq.(\ref{eq:Q}) equal to unity, then the naive evaluation of energies yield

\begin{equation}
E^2=J_1+J_2=E_1^2+E_2^2 \leq (E_1+E_2)^2=E^{'2}   \label{eq:R}
\end{equation}
where $E$ and $E'$ denote the energies corresponding to figures (a) or (c). In the case of figure (b), $J_1$, $J_2$ are the angular momenta corresponding to the energies $E_1$ and $E_2$ of the subsystems. The equality in Eq.(\ref{eq:R}) holds only if $E_1$ or $E_2$ is zero. Therefore for each fixed total angular momentum its most unfair partition into two subsystems gives us the lowest energy levels, and its more or less fair partition gives rise to energy levels on daughter trajectories. Hence on the leading baryonic trajectory we have a quark-diquark structure (Fig.(a)), or a linear molecule structure (Fig.(c)). On the other hand on low-lying trajectories we have more or less symmetric ($J_1 \sim J_2$) configuration of quarks.

Since high $J$ hadronic states on leading Regge trajectories tend to be bilocal with large separation of their constituents, they fulfill all the conditions for supersymmetry between $\bar{q}$ and $D$. Then the only difference between the energies of ($q\bar{q}$) mesons and ($qD$) baryons comes from the different masses of their constituents, namely $m_q=m_{\bar{q}}=m$, and $m_D\sim 2m$. For high $J$ this is the main source of symmetry breaking which is spin independent. We will show how we can obtain sum rules from this breaking. The part of the mass operator that gives rise to this splitting is a diagonal element of $U(6/21)$ that commutes with $SU(6)$. 

Let us now consider the spin dependent breaking of $U(6/21)$. For low $J$ states the ($qD$) system becomes trilocal($qqq$), the flux tube degenerates to a single gluon propagator that gives spin-dependent forces in addition to the Coulomb term $ \frac{a}{r} $. In this case we have the regime studied by de Rujula, Georgi and Glashow, where the breaking is due to hyperfine splitting caused by the exchange of single gluons that have spin $1$. These mass splittings give rise to different intercept of the Regge trajectories are given by

\begin{equation}
\Delta m_{12}=k \frac{ \bf{{\vec{ S}_1} \cdot {\vec{S}_2 }}}{m_1m_2},~~~~k=|\psi(0)|^2
\end{equation}
both for baryons and mesons at high energies. But at low energies the baryon becomes a trilocal object (with three quarks) and the mass splitting is given by$^{\cite{kac}}$

\begin{equation}
\Delta m_{123}= \frac{1}{2} k(\frac{ \bf{{\vec{ S}_1} \cdot {\vec{S}_2 }}}{m_1m_2}+\frac{ \bf{{\vec{ S}_2} \cdot {\vec{S}_3 }}}{m_2m_3}+
\frac{ \bf{{\vec{ S}_3} \cdot {\vec{S}_1 }}}{m_3m_1}
\end{equation}
where $m_1$, $m_2$ and $m_3$ are the masses of the three different quark constituents. 

The element of $U(6/21)$ that give rise to such splittings is a diagonal element of its $U(21)$ subgroup and gives rise to $s(s+1)$ terms that behave like an element of the (405) representation of $SU(6)$ in the $SU(6)$ mass formulae. The splitting of isospin multiplets is due to a symmetry breaking element in the (35) representation of $SU(6)$. hence all symmetry breaking terms are in the adjoint representation of $U(6/21)$. If we restrict ourselves to the non-strange sector of hadrons with approximate $SU(4)$ symmetry, effective supersymmetry will relate the splitting in $m^2$ between the $\Delta$ ($s=\frac{3}{2}, I=\frac{3}{2}$) and $N$ ($s=\frac{1}{2}, I=\frac{1}{2} $) to the splitting between $\omega$ ($s=1, I=0$) and $\pi$ ($s=0, I=1$) so that

\begin{equation}
m_\Delta^2 -m_N^2 \simeq m_\omega^2 - m_\pi^2
\end{equation}
which is satisfied to within 5\%.

\begin{center}
{ \bf  Effective Hamiltonian of the Relativistic Quark Model} 
\end{center}                             
                              
The quark model is considered to be an effective theory of QCD at low                                
energies when the gluon degrees of freedom are integrated out.                                
A relativistic potential model of quarks can be built based upon the effective                                
color potentials.                                
Lattice gauge calculations indicate that, inside a hadron, quarks are bounded                                
by a confining potential that grows like $r$, where $r$ is the separation                                
between two quarks, for large $r$. At small $r$, the color force is                                
coulomb-like, i.e. $\sim 1/r$.                                
                                
Following the treatment given in our previous papers$^{ \cite{cg1}, \cite{cg2}}$ we                                
consider a quark $q(1)$ at position $\mbox{\boldmath $r$}_{1}$, and an                                
antiquark $\overline{q}(2)$ at                                
position $\mbox{\boldmath $r$}_{2}$ represented by the charge                                
conjugate spinor $q^{c}(2)$. The $q(1)\overline{q}(2)$ system for a meson has                                
$16$ space components that are the elements of the $4 \times 4$ matrix                                

\begin{equation}                                
\Phi(1,2)=q(1)\overline{q}(2).                   
\end{equation}                                
Elimination of gluon degrees of freedom generates a potential between these                                
constituents that in the static approximation depends on the relative distance                                
$r$. It will be approximated by a scalar potential                                
$S(r)$ and the fourth component of a vector potential $V(r)$ given by                                

\begin{equation}                                
S(r)=br,                                
\end{equation}                                

\begin{equation}                                
V(r)=-\frac{4}{3}\alpha_{s}\frac{1}{r},                                
\end{equation}                                
where $\alpha_{s}$ is the strong coupling constant at an energy scale                                
corresponding to low lying meson masses. In this approximation the relativistic                                
Hamiltonian can be written as                                

\begin{equation}                                
H\Phi=(H_{(1)}+H_{(2)})\Phi,                                
\end{equation}                                
with                                

\begin{equation}                                
H_{(i)}=\frac{1}{2}V(r)+\gamma^{(i)}_{4}[m_{i}+\frac{1}{2}S(r)]                                
+i\gamma^{(i)}_{4}\mbox{\boldmath $\gamma$}^{(i)}                                
\mbox{\boldmath $\cdot p$}^{(i)},                                
\end{equation}                                
where $i=1,2$, and $\mbox{\boldmath $p^{(i)}$}                                
=-i\mbox{\boldmath $\nabla$}^{(i)}$.                                
At this point let us introduce the center of mass and relative coordinates

\begin{equation}
\mbox{\boldmath $R$} = \frac{m_{1} \mbox{\boldmath $r$}_{1} +
m_{2} \mbox{\boldmath $r$}_{2}}{m_{1} + m_{2}}, ~~~~
                    \mbox{\boldmath $r$} =
\mbox{\boldmath $r$}_{1} - \mbox{\boldmath $r$}_{2},
\end{equation}

\begin{equation}
\mbox{\boldmath $r$}_{1} = \mbox{\boldmath $R$} + \frac{m_{2}}{m_{1}+m_{2}}
\mbox{\boldmath $r$}, ~~~~
\mbox{\boldmath $r$}_{2} = \mbox{\boldmath $R$} - \frac{m_{1}}{m_{1}+m_{2}}
\mbox{\boldmath $r$}, 
\end{equation}
we have

\begin{equation}
\mbox{\boldmath $p$}_{1} = m_{1} \dot{\mbox{\boldmath $R$}} +
\mu \dot{\mbox{\boldmath $r$}}, ~~~~
\mbox{\boldmath $p$}_{2} = m_{2} \dot{\mbox{\boldmath $R$}} -
\mu \dot{\mbox{\boldmath $r$}} ,
\end{equation}
where $\mu$ is the reduced mass.                                
In the center of mass system $({\bf R}=0)$, we find                             

\begin{equation}                              
\mbox{\boldmath $p^{(1)}$} + \mbox{\boldmath $p^{(2)}$} =0,                              
\end{equation}                              
\begin{equation}                              
\mbox{\boldmath $p^{(1)}$} = - \mbox{\boldmath $p^{(2)}$} =                              
\mbox{\boldmath $p$} = -i \mbox{\boldmath $\nabla$} ,                              
\end{equation}                              
and                              

\begin{equation}                              
\mbox{\boldmath $r$} =                              
\mbox{\boldmath $r^{(1)}$} - \mbox{\boldmath $r^{(2)}$} .                              
\end{equation}                              
                              
Since $\gamma_{\mu}^{(1)}$ and $\gamma_{\mu}^{(2)}$ commute we can represent                                
them by left and right multiplication of $\Phi$ by Dirac matrices so that                                

\begin{equation}                                
\gamma_{\mu}^{(1)}\Phi=\gamma_{\mu}\Phi,                                
\;~~~~~~ \gamma_{\mu}^{(2)}\Phi=\Phi\gamma_{\mu}.                                
\end{equation}                                
                              
We now make a unitary non-local transformation similar to the Foldy-Wouthuysen                              
transformation by defining a new two-body wave function                                

\begin{equation}                                
\tilde{\Phi}=W\Phi=W_{1}W_{2}\Phi,                                
\end{equation}                                
where                                

\begin{equation}                                
W_{i}=\left[\frac{m_{i}+S/2+i\mbox{\boldmath $\gamma$}^{(i)}                                
\mbox{\boldmath $\cdot p$}^{(i)}}                                
{\sqrt{(m_{i}+S/2)^{2}+\mbox{\boldmath $p$}^{(i)}                                
\mbox{\boldmath $\cdot p$}^{(i)}}}\right]^{1/2}                                
\end{equation}                                
and $W_{i}W_{i}^{\dagger}=1$ for $i=1,2$.                                 
Since $\vec{x}$ does not commute with                                 
$\mbox{\boldmath $p$}^{(i)}$ the exact                                
form of the transformed Hamiltonian $\tilde{H}$ is very complicated. Instead                                
we use an approximation for W by replacing S by an average value                                
\begin{equation}                                
<S> = S(r_{0})                                
\end{equation}                                
where $r_{0}$ is the semiclassical radius of the Bohr orbit associated with                                 
the $q-\bar{q}$ potential. In this simplified version W depends only on                                 
differential operators, so that we can calculate $\tilde{H}$ by expanding                                
W in the powers of the momenta. $\tilde{H}$ will also involve a transformed                                 
vector potential $\tilde{V}(r)$,                                

\begin{equation}                                
\tilde{H}=WHW^{\dagger}, \;~~~~~~ \tilde{V}(r)=WV(r)W^{\dagger}.                                
\end{equation}                                
In the approximation that $\tilde{S}(r)$ is spin-independent, we have                                

\begin{equation}                                
\tilde{S}(r)=WS(r)W^{\dagger}\approx S(r)=S(r_{0}) +((S(r)-S(r_{0})),                                
\end{equation}                                

\begin{equation}                                
\tilde{V}(r)\approx V(r)+K\frac{\mbox{\boldmath $\sigma$}^{(1)}                                
\mbox{\boldmath $\cdot\sigma$}^{(2)}}{4m_{1}m_{2}}\nabla^{2}V(r),                                
\end{equation}                                
so that $\tilde{V}(r)$ differs from $V(r)$ by a spin-dependent                               
Breit term.                               
                              
The new two-body equation now reads                                

\begin{equation}                                
\tilde{H}\tilde{\Phi}=\tilde{V}(r)\tilde{\Phi}+                                
\sqrt{[m_{1}+\frac{1}{2}S(r)]^{2}+\mbox{\boldmath $p$}_{(1)}^{2}}\gamma_{4}                                
\tilde{\Phi}                                
+\sqrt{[m_{2}+\frac{1}{2}S(r)]^{2}+\mbox{\boldmath $p$}_{(2)}^{2}}\tilde{\Phi}                                
\gamma_{4}.             \label{eq:ener}                              
\end{equation}                                
                              
This equation is invariant under the spin transformation                              
                              
\begin{equation}                              
\tilde{\Phi} \rightarrow  e^{\frac{1}{2} i                               
\mbox{\boldmath $\sigma$}                                
\mbox{\boldmath $\cdot\omega_{1}$}} \Phi                              
e^{\frac{-1}{2} i                               
\mbox{\boldmath $\sigma$}                                
\mbox{\boldmath $\cdot\omega_{2}$}}                              
\end{equation}                              
since                              

\begin{equation}                              
[\mbox{\boldmath $\sigma$}, \gamma_{4}] = 0,                              
\end{equation}                              
provided the Breit term is neglected. The potential being flavor-independent,                              
flavor independence is only broken by the flavor-dependent mass differences                              
$m_{1}-m_{2}$. This leads to an approximate $SU(6) \times SU(6)$ symmetry                              
of $\tilde{H}$ broken by mass differences and spin dependent Breit terms.                              
Note that the original Hamiltonian $H$ does not display this invariance. The                              
invariance of $\tilde{H}$ does not conflict with relativity since it is                              
obtained by a non-local transformation.                              
                              
Iteration of Eq.(\ref{eq:ener}) gives the "Krolikowski" type second                              
order equation
                               
\begin{equation}                              
-\nabla^{2} \tilde{\Phi} = \frac{1}{4}                               
[1- (\frac{m_{1}-m_{2}}{E-\tilde{V}(r)})^{2}]                              
[(E-\tilde{V}(r))^{2}-(m_{1}+m_{2}+S(r))^{2}] \tilde{\Phi}                              
\end{equation}                              
which was successfully used by Lichtenberg and his collaborators$^{ \cite{l1}} $                              
to calculate meson masses in the approximation $\tilde{V}=V$.                              
                              
For a $q-D$ system $\bar{q}(2)$ is replaced by a wave function with four spin                              
components associated with $s=1$ and $s=0$. In addition there are                               
flavor quantum numbers. We write                              

\begin{equation}                              
\tilde{V}(r) = V(r) + K \frac{\mbox{\boldmath $S$}^{(1)}                                
\mbox{\boldmath $\cdot S$}^{(2)}}{m_{1}m_{2}} \nabla^{2} V(r),                                
\end{equation}                                
with $m_{1}$ and $m_{2}$ being respectively the quark and diquark masses,                              
$\boldmath{S}^{(1)}$ the $s=1/2$ Pauli matrices and $\boldmath{S}^{(2)}$                              
being either zero or the $3 \times 3$ s=1 matrices. Otherwise the two-body                              
equation is the same as for the meson case. If $m_{1}-m_{2}$ and the                               
spin-dependent terms are neglected the two-body equation for the                               
meson-baryon system in which the wave function is                              

\begin{equation}                                
\tilde{\Psi}=\left(                                
\begin{array}{cc}                                
\tilde{\Phi}_{q\overline{q}} & \tilde{\Psi}_{qD} \\                                
\tilde{\Psi}_{\overline{D}\overline{q}} & \tilde{\Phi}_{\overline{D}D}                                
\end{array} \right)                              
\end{equation}                                
becomes invariant under the transformation                              

\begin{equation}                              
\tilde{\Psi} \rightarrow U \tilde{\Psi} V^{\dag}                              
\end{equation}                              
where $U$ and $V$ are elements of the Miyazawa supergroup $SU(6/21)$. In a                               
relativistic treatment we can represent the diquark for each flavor by a                              
$16$-component wave function satisfying a linear Kemmer type equation. $10$                              
components are for $s=1$  ($F_{\mu \nu}$ and $A_{\lambda}$) and $5$ are for $s=0$                              
($\phi_{\mu}$ and $\phi$). Adding $4$ Dirac components for the quark with                              
$s=1$, we find that for each flavor state the $q-\bar{D}$ system has $20$                              
covariant components representing $6$ physical spin states                              
$s$ =$\frac{1}{2}$, $0$ and $1$.                              
                              
The matrix wave function $\tilde{\Psi}$ obtained from the covariant wave                               
function after a non-local transformation has the physical states                               
separated in a non-covariant way expressible in terms of $27 \times 27$ matrix.                              
It satisfies a wave equation of the form                              

\begin{equation}                              
H \tilde{\Psi} = i \frac{\partial \tilde{\Psi}}{\partial t} =\{K, \tilde{\Psi}\}                              
+ \mbox{\boldmath $S$} \tilde{\Psi} \mbox{\boldmath $S$}     \label{eq:K}                              
\end{equation}                              
where                              

\begin{equation}                                
 K =\left(                                
\begin{array}{cc}                                
K_{q}I^{(6)} & 0 \\                                
0 & K_{D} I^{(21)}                                
\end{array} \right)                                       
\end{equation}                              
with $I^{(n)}$ denoting the $n \times n$ unit matrix and                              

\begin{equation}                              
K_{q}= -\frac{2}{3} \frac{\alpha_{s}}{r} + \gamma_{4}                              
[(m_{q} + \frac{1}{2} S)^{2} -\nabla^{2}]^{\frac{1}{2}} ,                              
\end{equation}                                

\begin{equation}                              
K_{D}= -\frac{2}{3} \frac{\alpha_{s}}{r} + \beta_{4}                              
[(m_{q} + \frac{1}{2} S)^{2} -\nabla^{2}]^{\frac{1}{2}} ,                              
\end{equation}                                

\begin{equation}                              
\mbox{\boldmath $S$}= K^{\frac{1}{2}}                               
\left(                                
\begin{array}{cc}                                
m_{q}^{-1} \mbox{\boldmath$S$}_{q} I^{(6)} & 0 \\                                
0 & m_{D}^{-1} \mbox{\boldmath$S$}_{D} I^{(21)}                              
\end{array} \right)                              
\end{equation}                              
where $\beta_{4}$ is the fourth Petiau-Kemmer matrix for $s=0$ or $1$ can be                              
represented by                               

\begin{equation}                              
\beta_{4} = \gamma_{4}^{(2)} + \gamma_{4}^{(3)}                              
\end{equation}                              
referring to the two quark                              
constituents of the diquark. $\mbox{\boldmath S}_{q}$ and                               
$\mbox{\boldmath S}_{D}$                              
are respectively the quark and diquark spin operators, $m_{q}$ and                               
$m_{D}$ are their respective masses. The second term in Eq.(\ref{eq:K}) gives                               
the spin dependent Breit term and breaks the approximate $SU(6/21)$                              
symmetry generated by                              

\begin{equation}                              
\delta \tilde{\Psi} = [X, \tilde{\Psi}],                              
\end{equation}                              
\begin{equation}                              
X =                               
\left(                                
\begin{array}{cc}                                
M & B \\                                
\bar{B} & N                              
\end{array} \right)                              
\end{equation}                              
where $B$ and $\bar{B}$ are Grassmann valued, $M$ and $N$ are antihermitian $6 \times 6$                              
and $21 \times 21$ matrices respectively. $S$ is the confining scalar potential. We note that                              

\begin{equation}                              
- \nabla^{2} = \mbox{\boldmath $p$}^{2} = p_{r}^{2} +                               
\frac{\ell(\ell+1)}{r^{2}} ,                              
\end{equation}                              
where $p_{r}$ is canonically conjugate to r.

The ground state energy eigenvalue $E$ of the Hamiltonian                              
can be estimated by using the Heisenberg uncertainty principle. This leads                              
to the replacement of $r$ by $\Delta r$ and $p_{r}$ by                              

\begin{equation}                              
\Delta p_{r} = \frac{1}{2} (\Delta r)^{-1}, ~~~~~~(h = 1).                              
\end{equation}                              
                              
Then $E$ as a function of $\Delta r$ is minimized for the value of                              
$r_{0}$ of $\Delta r$. The $r_{0}$ corresponds to the Bohr radius for                              
the bound state. The confining energy associated with this Bohr radius                               
is obtained from the linear confining potential                              

\begin{equation}                              
S(r) = br.                              
\end{equation}                              
so that the effective masses of the constituents become                              

\begin{equation}                              
M_{1} = m_{1} + \frac{1}{2} S_{0}, ~~~~                              
M_{2} = m_{2} + \frac{1}{2} S_{0}                              
\end{equation}                              

\begin{equation}                              
S_{0} = b r_{0}.                              
\end{equation}                              
                              
For a meson $m_{1}$ and $m_{2}$ are the current quark masses while                              
$M_{1}$ and $M_{2}$ can be interpreted as the constituent quark                                
masses. Note that even in the case of vanishing quark masses associated                              
with perfect chiral symmetry, confinement results in non zero                              
constituent masses that spontaneously break the $SU(2) \times SU(2)$                               
symmetry of the $u$, $d$ quarks.                              
                              
Let us illustrate this method on the simplified spin free Hamiltonian                              
involving only the scalar potential. In the center of mass system,                                
$\mbox{\boldmath $p$}^{(1)}+\mbox{\boldmath $p$}^{(2)}=0$, or                                
$\mbox{\boldmath $p$}^{(1)}=-\mbox{\boldmath $p$}^{(2)}=\mbox{\boldmath $p$}$.                                
The semi-relativistic Hamiltonian of the system is then given by                                

\begin{equation}                                
E_{12}~~ \Phi=\sum_{i=1}^{2}\sqrt{(m_{i}+\frac{1}{2}br)^{2}+                              
\mbox{\boldmath $p$}^{2}}~~\Phi.                                
\end{equation}                                
Taking $m_{1}=m_{2}=m$ for the quark-antiquark system, we have                                

\begin{equation}                                
E_{12}~~\Phi= 2\sqrt{(m+\frac{1}{2}br)^{2}+p^{2}_{r}+                              
\frac{\ell(\ell+1)}{r^{2}}} ~~\Phi,           \label{eq:e}                                
\end{equation}                                
where we have written the momentum part in spherical coordinates.                              
                              
Putting                              

\begin{equation}                              
b=\mu^{2},~~~~~\rho = \mu ~~r,                              
\end{equation}                              
for the $q-\bar{q}$ system we find $E_{12}$ by minimizing the function                              

\begin{equation}                                
E_{q \bar{q}}~~= 2\sqrt{(m+\frac{1}{2} \mu \rho)^{2}+                              
\frac{\mu^{2}}{\rho^{2}} (\ell+\frac{1}{2})^{2}} .                                
\end{equation}                                
                                
For $u$ and $d$ quarks, $m$ is small and can be neglected so that                                

\begin{equation}                              
E^{2} = \mu^{2} [\rho^{2} + \rho^{-2} (2 \ell + 1)^{2}]
\end{equation}                              
which has a minimum for                              

\begin{equation}                              
\rho^{2} = \rho_{0}^{2} = 2 \ell + 1,                              
\end{equation}                              
giving                              

\begin{equation}                              
E_{min}^{2} = E^{2} (\rho_{0}) = 4 \mu^{2} (\ell+\frac{1}{2}).                       \label{eq:min}       
\end{equation}                              
Thus, we obtain a linear Regge trajectory with                              

\begin{equation}                              
\alpha^{'} = \frac{1}{4} \mu^{-2} = \frac{b}{4}.                       \label{eq:mina}       
\end{equation}                              
Also                              

\begin{equation}                              
\mbox{\boldmath $J$}= \mbox{\boldmath $\ell$} +\mbox{\boldmath $S$}                              
\end{equation}                              
where $\mbox{\boldmath $S$}$ arises from the quark spins. Experimentally                              

\begin{equation}                              
\alpha^{'} = 0.88 (GeV)^{-2}                              
\end{equation}                              
for mesons giving the value $0.54$ GeV for $\mu$. A more accurate calculation$^{\cite{egu},\cite{jo}} $ gives                              

\begin{equation}                              
\alpha^{'} = (2 \pi \mu^{2})^{-1}, ~~~~~ \mu \sim 0.43 GeV.   \label{eq:alp}                              
\end{equation}      
                              
The constituent quark mass can be defined in two ways                              

\begin{equation}                              
M_{c}(\ell) = \frac{1}{2} E_{min} = \mu \sqrt{\ell + \frac{1}{2}} ,                              
\label{eq:mc}                              
\end{equation}                              
or                              

\begin{equation}                              
m_{c}^{'}(\ell)=S_{0} = \frac{1}{2} \mu \rho_{0} =                               
\frac{\mu}{\sqrt{2}} \sqrt{\ell + \frac{1}{2}}.                    \label{eq:mca}          
\end{equation}                              
The first definition gives for $\ell = 0$,                              

\begin{equation}                              
M_{c} = 0.31 GeV ~~~for~~~\mu= 0.43    \label{eq:mcb}                              
\end{equation}                              
in the case of $u$ and $d$ quarks.                              
                              
When the Coulomb like terms are introduced in the simplified Hamiltonian                              
Eq.(\ref{eq:e}) with negligible quark masses one obtains                              

\begin{equation}                              
E=\frac{\mu}{\rho} [-\bar{\alpha}  + \sqrt{\rho^{4} +      
(2\ell + 1)^{2}}]                              
\end{equation}                              
with                              

\begin{equation}                              
\bar{\alpha} =\frac{4}{3} \alpha_{s} ~~~{\rm for}~~ (q\bar{q}),~~~                              
\bar{\alpha} =\frac{2}{3} \alpha_{s} ~~~{\rm for}~~ (q q).                              
\end{equation}                              
In the energy range around $1$ GeV $\alpha_{s}$ is of order of unity.                              
Estimates range from $0.3$ to $3$. Minimization of $E$ gives                              

\begin{equation}                              
E_{0}=\mu u_{0}^{\frac{-1}{4}} (-\bar{\alpha} + \sqrt{u_{0} +(2\ell + 1)^{2}})                              
\end{equation}                              
where                              

\begin{equation}                              
u_{0}(\epsilon)=\rho_{0}^{4}= (2 \ell + 1)^{2} (1+\frac{1}{2} \beta^{2} +                              
\epsilon \sqrt{2} \beta                              
\sqrt{\ell+\frac{1}{8} \beta^{2}}),                              
\end{equation}                              

\begin{equation}                              
\epsilon = \pm{1},~~~~\beta=\frac{\bar{\alpha}}{(2 \ell + 1)}.                              
\end{equation}                              
                              
The minimum $E_{0}$ is obtained for $\epsilon = -1$, giving to second                              
order in $\beta$:                              

\begin{equation}                              
E_{0} = \mu \sqrt{2(2 \ell + 1 )} (1 - \frac{\beta}{\sqrt{2}} -                              
3 \frac{\beta^{2}}{8}).       \label{eq:ez}                              
\end{equation}                              
                              
Linear Regge trajectories are obtained if $\beta^{2}$ is negligible.                              
Then for mesons                              

\begin{equation}                              
E_{0}^{2} = 4 \mu^{2} \ell + 2 \mu^{2} (1-\sqrt{2} \bar{\alpha}).                              
\end{equation}                              
The $\beta^{2}$ is negligible for small $\ell$ only if we take the lowest                              
estimate for $\alpha_{s}$, giving $0.4$ for $\bar{\alpha}$ in the $q \bar{q}$                              
case. For mesons with $u$, $d$ constituents, incorporating their spins through                              
the Breit term we obtain approximately                              

\begin{equation}                              
m_{\rho} \simeq m_{\omega} = E_{0} + \frac{c}{4}, ~~~~~                              
m_{\pi}= E_{0} - \frac{3c}{4},                              
\end{equation}                              

\begin{equation}                              
c = K~~ \frac{\Delta V}{M_{q}^{2}}                              
\end{equation}                              
where $M_{q}$ is the constituent quark mass. This gives                              

\begin{equation}                              
E_{0}= \frac{(3 m_{\rho} + m_{\pi})}{4} = 0.61 GeV.                               
\end{equation}                              
                              
The Regge slope being of the order of $1 GeV$ an average meson mass                              
of the same order is obtained from Eq.(\ref{eq:ez}) in the linear 
trajectory approximation. To this approximation $\bar{\alpha}$ should be treated like                               
a parameter rather than be placed by its value derived from QCD under                               
varying assumptions. Using the value shown in Eq.(\ref{eq:alp}) for $\mu$, one gets a better                               
fit to the meson masses by taking $\alpha_{s} \sim 0.2$.                              
                              
Turning now to baryon masses, we must first estimate the diquark mass. We                              
have for the $qq$ system                              

\begin{equation}                              
M_{D}= \mu (\sqrt{2} - \frac{2}{3} \alpha_{s}),                \label{eq:dq}           
\end{equation}                              
that is slightly higher than the average meson mass                              

\begin{equation}                              
\tilde{m}= \mu (\sqrt{2} - \frac{4}{3} \alpha_{s})            \label{eq:con}      
\end{equation}                              
                              
Here we note that $E$ is not very sensitive to the precise value of the QCD running coupling constant in the $GeV$ range. Taking $\alpha_{s} \sim 0.3$ changes $E^{qq}$ from $0.55$ to $0.56 GeV$.      
      
Note that Eq.(\ref{eq:dq}) gives $M_{D} = 0.55 GeV$. For excited $q-\bar{q}$ and $q-D$ systems if the rotational excitation energy is large compared with $\mu$, then both the $m_{D}$ and the Coulomb term $- \frac{4}{3} \frac{\alpha_{s}}{r}$ (same for $q-D$ and $q-\bar{q}$ systems) can be neglected.      
      
Thus, for both ($q-D$) [excited baryon] and $q-\bar{q}$ [excited meson] systems we have Eq.(\ref{eq:min}), namely      

\begin{equation}      
(E^{q-D})^{2} \sim (E^{q-\bar{q}})^{2} \sim 4 \mu^{2} \ell + 2 \mu^{2}      
\end{equation}      
giving again Eq.(\ref{eq:mina}), i.e.      

\begin{equation}      
(\alpha^{'})_{q-D} = (\alpha^{'})_{q-\bar{q}} \cong \frac{1}{4 \mu^{2}}~~~~{\rm or} ~~(\frac{1}{2 \pi \mu^{2}})      
\end{equation}      
as an explanation of hadronic supersymmetry in the nucleon and meson Regge spectra. We also have, extrapolating to small $\ell$:      

\begin{equation}      
\Delta (M^{2})^{q-D} = \Delta (m^{2})^{q-\bar{q}} = 4 \mu^{2} \Delta \ell = \frac{1}{\alpha^{'}} \Delta \ell .      
\end{equation}        
For $\Delta \ell = 1$ we find      

\begin{equation}      
m_{\Delta}^{2} - m_{N}^{2} = m_{\rho}^{2} - m_{\pi}^{2} .      
\end{equation}      
      
This relationship is same as the one proved by Miyazawa through the assumption that $U(6/21)$ symmetry is broken by an operator that behaves like $s = 0$, $I = 0$ member of $35 \times 35$ representations of $SU(6)$, which is true to $5\%$. It corresponds to a confined quark approximation with $\alpha_{s} = 0$.      
      
The potential model gives a more accurate symmetry breaking ($\alpha_{s} \sim 0.2$):      

\begin{equation}      
\frac{9}{8} (m_{\rho}^{2} - m_{\pi}^{2}) = m_{\Delta}^{2} - m_{N}^{2}      
\end{equation}      
with an accuracy of $1\%$, in a remarkable agreement with experiments.       
      
This mass squared formula arises from the second order iteration of the $q-D$, $q-\bar{q}$ Dirac equation. The factor $\frac{9}{8}$ comes from      

\begin{equation}      
\frac{1}{2} (\frac{4}{3} \alpha_{s})^{2} = \frac{8}{9} \alpha_{s}^{2} .      
\end{equation}      
      
At this point it is more instructive to derive a first order mass formula.

Since the constituent quark mass $M_{q}$ is given by Eq.(\ref{eq:mc})                              
$(\ell = 0)$, we have                              

\begin{equation}                              
M_{q} = \frac{\mu}{\sqrt{2}},                              
\end{equation}                              
so that                              

\begin{equation}                              
\bar{m} = 2 M_{q} ( 1 -\frac{\sqrt{2}}{3} \alpha_{s}) \simeq 1.9 M_{q} .                              
\end{equation}                              
                              
When the baryon is regarded as a $q-D$ system, each constituent gains an                               
effective mass $\frac{1}{2} \mu \rho_{0}$ which was approximately the                              
effective mass of the quark in the meson. Hence, the effective masses of                              
$q$ and $D$ in the baryon are                               

\begin{equation}                              
m_{q}^{'} \simeq M_{q}, ~~~~~m_{D}^{'} = M_{D} + M_{q} \simeq 3 M_{q}.                              
\end{equation}                              
                              
The spin splittings for the nucleon $N$ and the $\Delta$ are given by                              
the Breit term                              

\begin{equation}                              
\Delta M = K \Delta V \frac{\mbox{\boldmath $S$}_{q}                                
\mbox{\boldmath $\cdot S$}_{D}}{m_{q}^{'} m_{D}^{'}}.                              
\end{equation}                              
                              
For the nucleon with spin $\frac{1}{2}$ the term ${\bf S}_{q} \cdot {\bf S}_{D}$                              
gives $-1$ while it has the value $\frac{1}{2}$ for $\Delta$ with spin $\frac{3}{2}$. Using the same $K$ for mesons and baryons which are                              
both considered to be a bound state of a color triplet with a color                               
antitriplet we can relate the baryon splitting $\Delta M$ to the meson                              
splitting $\Delta m$ for which ${\bf S}_{q} \cdot {\bf S}_{\bar{q}}$ takes the                              
values $\frac{1}{4}$ and $\frac{-3}{4}$. Hence we find                              

\begin{equation}                              
\Delta M = M_{\Delta} - M_{N} = \frac{3}{2} \frac{K \Delta V}                              
{m_{q}^{'} m_{D}^{'}} = \frac{3}{2} \frac{K \Delta V}                              
{3 M_{q}^{2}}                              
\end{equation}                              

\begin{equation}                              
\Delta m = \frac{K \Delta V}{M_{q}^{2}}                              
\end{equation}                              
which leads to a linear mass formula                              

\begin{equation}                              
\Delta M = \frac{1}{2} \Delta m                              
\end{equation}                              
which is well satisfied and has been verified before using the three                               
quark constituents for the baryon.                              
                              
The formation of diquarks which behave like antiquarks as far as QCD is                               
concerned is crucial to hadronic supersymmetry and to quark dynamics for                               
excited hadrons. The splittings in the mass spectrum are well understood                              
on the basis of spin-dependent terms derived from QCD. This approach to                               
hadronic physics has led to many in depth investigations recently. Examples                              
are the work of Martin$^{ \cite{ma}}$, Yamagishi$^{ \cite{yam}}$ and others, and more recently Klempt$^{\cite{klempt}}$, and Lichtenberg and
collaborators$^{\cite{l1}}$ and references contained in their review paper$^{\cite{anse}}$.

Let us now consider a quark-antiquark system in the approximation that the                                
potential is only a scalar.                                
In the center of mass system,                                
$\mbox{\boldmath $p$}^{(1)}+\mbox{\boldmath $p$}^{(2)}=0$, or                                
$\mbox{\boldmath $p$}^{(1)}=-\mbox{\boldmath $p$}^{(2)}=\mbox{\boldmath $p$}$.                                
The semi-relativistic Hamiltonian of the system is then given by                                

\begin{equation}                                
H=\sum_{i=1}^{2}\sqrt{(m_{i}+\frac{1}{2}br)^{2}+\mbox{\boldmath $p$}^{2}}.                                
\end{equation}                                
Taking $m_{1}=m_{2}=m$ for the quark-antiquark system, we have                                

\begin{equation}                                
H=2\sqrt{(m+\frac{1}{2}br)^{2}+p^{2}_{r}+\frac{\ell(\ell+1)}{r^{2}}},                                
\end{equation}                                
where we have written the momentum part in spherical coordinates.                                
For $u$ and $d$ quarks, $m$ is small and can be neglected so that                                

\begin{equation}                                
H^{2}=4[\frac{1}{4}b^{2}r^{2}+p^{2}_{r}+\frac{\ell(\ell+1)}{r^{2}}].                                
\end{equation}                                
Similar equations for a $q-\bar{q}$ meson system were already proposed                                 
and solved numerically$^{\cite{l1}}$ or exactly$^{ \cite{g3}}$ by several authors.                                
The eigenfunction $\Psi$ for $H^{2}$ has eigenvalue $E^{2}$ such that                                

\begin{equation}                                
4[\frac{1}{4}b^{2}r^{2}-\frac{1}{r}\frac{d^{2}}{dr^{2}}r                                
+\frac{\ell(\ell+1)}{r^{2}}]\Psi=E^{2}\Psi.                                
\end{equation}                                
The differential equation can be solved exactly, and                                
and the normalized eigenfunction is found to be                                

\begin{eqnarray}                                
\Psi(r,\theta,\phi)&=&\left[\frac{2(\frac{b}{2})^{2\ell+3}                                
\Gamma(\ell+\frac{3}{2}+n_{r}-1)}                                
{(n_{r}-1)!\Gamma^{2}(\ell+\frac{3}{2})}\right]^{1/2}                                
r^{\ell}\exp[-\frac{b}{4}r^{2}]\times  \nonumber \\                                
&&F(-n_{r}+1,{\textstyle \ell+\frac{3}{2},\frac{b}{2}r^{2}})                                
Y_{\ell}^{m}(\theta,\phi),                                
\end{eqnarray}                                
where $F(-n_{r}+1,\ell+\frac{3}{2},\frac{b}{2}r^{2})$ is the confluent                                
hypergeometric function, and $n_{r}=1,2,3,\ldots$, is the radial quantum                                
number. The eigenvalue is given by                                

\begin{equation}                                
E^{2}=4b[2(n_{r}-1)+\ell+\frac{3}{2}],  \label{eq:regge}                                
\end{equation}                                
therefore we obtain linear Regge trajectories of slope $1/4b$ when we make                                
plots of $\ell$ versus $M^{2}$.  The case of                                
$n_{r}=1$ corresponds to the leading Regge trajectory, and cases of                                
$n_{r}=2,3,\ldots$, correspond to the parallel daughter trajectories.

Iachello and his collaborators$^{ \cite{i1}}$ have obtained a similar mass                                
formula based on algebraic methods. Starting from a spectrum generating                                
algebra $G$, we can write a chain of subalgebras                                

\begin{equation}                                
G \supset G' \supset G'' \supset \cdots .                                
\end{equation}                                
They proposed that the Hamiltonian can be expanded in terms of                                
invariants of the chain of subalgebras                                

\begin{equation}                                
H=\alpha C(G)+\beta C(G')+\gamma C(G'')+\cdots,                                
\end{equation}                                
where $C(G)$ denotes one of the invariants of $G$. In a two-body problem, the                                
spectrum generating algebra is U(4), and in the total symmetric representation,                                
one of the two chains of subalgebras is U(3) followed by SO(3) with                                
respective Casimir                                
invariants $n(n+2)$ and $\ell(\ell+1)$, where $n$ corresponds to the                                
vibrational mode.  In order to derive the relativistic mass formula like                                
Eq.(\ref{eq:regge}), one has to write a formula for the square of the mass, rather                                
than the mass itself.  Furthermore, without violating the dynamic symmetry,                                
the Hamiltonian can be written in terms of non-linear functions of the Casimir                                
invariants. Therefore, we can have                                

\begin{eqnarray}                                
M^{2}&=&\alpha+\beta\sqrt{n(n+2)+1}+\gamma\sqrt{\ell(\ell+1)+1/4} \nonumber \\                                
&=&\alpha '+\beta n+\gamma \ell .                        \label{eq:iachello}                                
\end{eqnarray}                                
Comparing this mass formula with Eq.(\ref{eq:regge}), we find that the relativistic                                
quark model suggests a relation of $\beta=2\gamma$ in Eq.(\ref{eq:iachello}).                                
However, the above algebraic method treats $\beta$ and $\gamma$ as two                                
independent parameters, and so the extra degree of freedom allows                                
Eq.(\ref{eq:iachello}) a better fit to the experimental data. The fitted                                
parameters for mesons are found to be: $\beta\approx 1.5$ (GeV)$^{2}$ and                                
$\gamma\approx 1.1$ (GeV)$^{2}$.                                
                                
The inclusion of the Coulomb-like term causes deviations from the linear                                
trajectories at low energies and changes the relation between the parameters                                
$\beta$ and $\gamma$. It is important to carry out a calculation based on a                                
better approximation to see if the empirical values of these parameters are                                
compatible with our model.                                
                                
A meson is made up of a quark and an antiquark, while a baryon is made up of                                
three quarks.  The complicated three-body problem in a baryon can be simplified                                
to a two-body one if we assume two of the three constituent quarks form a                                
diquark system.  We have recently$^{ \cite{cg1},\cite{cg2}}$  discussed                                
a possible hadronic supersymmetry of baryons and mesons suggested by the                                
symmetry between a quark-diquark system and a quark-antiquark system.                                
                                
In the color group SU(3) of QCD, a quark is in the ({\bf 3})                                
representation, while an                                
antiquark is in the ($\overline{{\bf 3}}$) representation.  Since                                
${\bf 3}\otimes{\bf 3}={\bf 6}\oplus\overline{{\bf 3}}$, a diquark can be                                
in ({\bf 6}) or                                
($\overline{{\bf 3}}$).   The color force due to one gluon exchange is                                
repulsive in the ({\bf 6}) representation, attractive in                                
the ($\overline{{\bf 3}}$) representation, with                                
the attractive force between two quarks being half in strength as that                                
of between a quark and an antiquark.                                
Therefore, a diquark is energetically favored to be in                                
the ($\overline{{\bf 3}}$).  From the representations of the color group SU(3),                                
QCD suggests a symmetry between                                
an antiquark $\overline{q}$ and a diquark $D$, both of ($\overline{{\bf 3}}$)                                
representation, and similarly for a quark $q$ and an                                
antidiquark $\overline{D}$ in the ({\bf 3}) representation.                                
                                
In the spin-flavor SU(6) symmetry, $q$ and $D$ lie respectively in the                                
({\bf 6}) and ($\overline{{\bf 21}}$) representations of SU(6).                                
One can put a quark (fermion) and an antidiquark (boson) into a superymmetric                                
multiplet                                

\begin{equation}                                
\psi=\left( \begin{array}{c} q \\ \overline{D} \end{array} \right)                                
\end{equation}                                
The effective supersymmetry can be described by the graded algebra SU(6/21)                                
with 6 fermionic and 21 bosonic components. We obtain a 27$\times$27 hadron                                
multiplet by forming the matrix                                

\begin{equation}                                
\psi\overline{\psi}=\left(                                
\begin{array}{cc}                                
q\overline{q} & qD \\                                
\overline{D}\overline{q} & \overline{D}D                                
\end{array} \right).                                
\end{equation}                                
This scheme predicts the existence of exotic mesons $\overline{D}D$, as well as                                
mesons ($q\overline{q}$), baryons ($qD$) and antibaryons                                
($\overline{D}\overline{q}$).                                
                                
The supersymmetry produces naturally the result that baryons and mesons have the                                
same slope of their Regge trajectories.  This prediction agrees with                                
experiment at high angular momenta.  Experimentally, the slopes of the Regge                                
trajectories for mesons and baryons are both approximately 0.9 (GeV)$^{-2}$.                                
Therefore, the scalar potential $S(r)=br$ has the fitted parameter                                
$b\approx 0.3$ (GeV)$^{2}$.                                
                                
To explain the meson-baryon symmetry phenomenon, note that                                
at high angular momenta, a baryon resembles a                                
string-like object with a quark at one end and a diquark at the other.                                
The gluon flux tube                                
contracts to a very small cross-sectional area, and hence the two                                
quarks forming the                                
diquark must be very close together to make the diquark look like a single                                
point particle.                                
Neglecting the difference between the diquark and antiquark                                
masses, the baryon system becomes a meson-like system.                                
Both $D$ and $\overline{q}$ are in the ($\overline{{\bf 3}}$) color                                
representation, so that the $q$-$D$ potential is the same as the                                
$q$-$\overline{q}$ potential provided we neglect the spin dependence of the                                
color force.  The calculation of                                
the mass spectrum of mesons discussed earlier just applies to                                
baryons as well, so the Regge trajectories of mesons and baryons are                                
parallel. This remarkable manifestation of hadronic supersymmetry had found                                
no simple explanation based on QCD until the quark-diquark model was worked                                
out.                                
                                
The observed hadronic $SU(6/21)$ supersymmetry is not exact.  The breaking of this                                
supersymmetry has two origins.  First one is the $q$ and $D$ mass differences                                
as well                                
as mass differences among quarks.  This leads to different intercepts for                                
parallel Regge trajectories.  The second breaking comes from the                                
contribution to the potential from one gluon exchange. This potential is a                                
4-vector and is spin dependent. Since the quark and antiquark have spin $S=1/2$                                
and the diquark has $S=0$ or $S=1$, the spin dependent part of the $q$-$D$                                
potential is different from that of $q$-$\overline{q}$, causing supersymmetry                                
breaking.  Another effect is the deviation of the Regge trajectories from                                
linearity for low angular momenta, since the potential is no longer                                
proportional to the distance, and quark masses can no longer be neglected.                                
                                
To see the symmetry breaking effect, we consider a hadron that                                
approximates a two-body system. The quark, antiquark, and diquark will acquire                                
their effective masses under the influence of the effective potentials. There                                
are also spin-dependent interactions among the quarks.  Based on a                                 
semi-relativistic Hamiltonian for a quark-diquark system interacting with the                                
same potentials $\tilde{S}(r)$ and the $\tilde{V}(r)$ the mass of                                
a hadron will take the approximate form                                

\begin{equation}                                
m_{12}=m_{1}+m_{2}+K\frac{\mbox{\boldmath $S$}^{(1)}\cdot\mbox{\boldmath                                
$S$}^{(2)}}{m_{1}m_{2}},                                
\end{equation}                                
where $m_{i}$ and {\boldmath $S$}$^{(i)}$ $(i=1,2)$ are respectively                                
the constituent mass and the spin of a quark or a diquark.                                
The spin-dependent Breit term will                                
split the masses of hadrons of different spin values.                                
                                
If we assume                                

\begin{equation}                                
m_{q}=m_{\overline{q}}=m                                
\end{equation}                                
where $m$ is the constituent mass of $u$ or $d$ quarks, and                                
denote the mass of diquark as $m_{D}$, then this  approximation gives                                

\begin{equation}                                
m_{\pi}=(m_{q\overline{q}})_{S=0}=2m-K\frac{3}{4m^{2}},                                
\end{equation}                                

\begin{equation}                                
m_{\rho}=(m_{q\overline{q}})_{S=1}=2m+K\frac{1}{4m^{2}},                                
\end{equation}                                

\begin{equation}                                
m_{\Delta}=(m_{qD})_{S=3/2}=m+m_{D}+K\frac{1}{2mm_{D}},                                
\end{equation}                                

\begin{equation}                                
m_{N}=(m_{qD})_{S=1/2}=m+m_{D}-K\frac{1}{mm_{D}}.                                
\end{equation}                                
Eliminating $m$, $m_{D}$ and $K$, we obtain a mass relation                                

\begin{equation}                                
\frac{8}{3}\cdot\frac{2m_{\Delta}+m_{N}}{3m_{\rho}+m_{\pi}}=1+\frac{3}{2}\cdot                                
\frac{m_{\rho}-m_{\pi}}{m_{\Delta}-m_{N}},                                
\end{equation}                                
which agrees with experiment to 13\%.                                
                       
The supersymmetry based on the supergroup                       
$U(6/21)$ acts on a                        
quark and antidiquark situated at the same point $x_{1}$.  At the                       
point $x_{2}$ we can consider                        
the action of a supergroup with the same parameters, or one with                       
different parameters.  In the first case we have a global symmetry.  In the                        
second case, if we only deal with bilocal fields, the symmetry will be                        
represented by $U(6/21)\times U(6/21)$, doubling the Miyazawa supergroup.                        
On the other hand, if any number of points are considered, with                       
different parameters                        
attached to each point, we are led to introduce a local supersymmetry                       
$U(6/21)$ to which we should add the local color group $SU(3)^{c}$.                        
Since it is not a                        
fundamental symmetry, we shall not deal with the local Miyazawa group here.                         
However, the double Miyazawa supergroup is useful for bilocal fields since the                        
decomposition of the adjoint representation of the 728-dimensional Miyazawa                        
group with respect to $SU(6)\times SU(21)$ gives                       

\begin{equation}                      
728= (35,1)+(1,440)+(6,21)+ (\bar{6}, \bar{21}) +(1,1)                      
\end{equation}                       
                       
A further decomposition of the double Miyazawa supergroup into its field with                        
respect to its c.m. coordinates leads                        
to the decomposition of the 126-dimensional cosets $(6,21)$ and                       
$(21,6)$ into $56^{+} + 70^{-}$  of the diagonal $SU(6)$.                      
                       
   We would have a much tighter and more elegant scheme if we could perform                        
such a decomposition from the start and be able to identify the $(1,21)$                       
part of the fundamental representation of $U(6/21)$ with the $21$-dimensional                       
representation of the $SU(6)$ subgroup, which means going beyond the                       
Miyazawa supersymmetry to a                        
smaller supergroup having $SU(6)$ as a subgroup.  Next, we describe                      
the development of a bilocal treatment, and a new minimal scheme                       
where the bilocal treatment gets carried over unchanged into it.                        
                      
\begin{center}

{\bf  Bilocal approximation to hadronic structure and inclusion of color.}

\end{center}                       
                       
~~~~   Low-lying baryons occur in the symmetric $56$ representation$^{ \cite{gra}}$                       
of $SU(6)$,                        
whereas the Pauli principle would have led to the antisymmetrical $20$                        
representation.  This was a crucial fact for the introduction of color degree                        
of freedom$^{\cite{gre}}$ based on $SU(3)^{c}$. Since the quark field                       
transforms like a color triplet and the diquark like a color                       
antitriplet under $SU(3)^{c}$, the color                        
degrees of freedom of the constituents must be included correctly in order                        
to obtain a correct representation of the q-D system.  Hadronic states must                        
be color singlets.  These are represented by bilocal operators                       
$O({\bf r}_{1},{\bf r}_{2})$ in the bilocal 
approximation$^{\cite{tak}}$                       
that gives $\bar{q}(1)q(2)$ for mesons and $D(1) q(2)$ for                        
baryons.  Here $ \bar{q}(1)$ represents the antiquark situated at ${\bf r}_{1}$,                       
$q(2)$ the quark situated at ${\bf r}_{2}$, and $D(1) = q(1)q(1)$ the                       
diquark situated at ${\bf r}_{1}$.  If we denote                        
the c.m. and the relative coordinates of the consituents by ${\bf R}$ and                       
${\bf r}$, where ${\bf r} ={\bf r}_{2}-{\bf r}_{1}$ and                       

\begin{equation}                      
{\bf R} = \frac{(m_{1} {\bf r}_{1}+m_{2} {\bf r}_{2})}{(m_{1}+m_{2})}                      
\end{equation}                       
with $m_{1}$ and $m_{2}$ being their masses, we can                        
then write $O({\bf R},{\bf r})$ for the operator that creates hadrons                       
out of the vacuum.                         
The matrix element of this operator between the vacuum and the hadronic state                        
$h$ will be of the form                        

\begin{equation}                      
<h| O({\bf R},{\bf r}) |0>= \chi({\bf R}) \psi({\bf r})                      
\end{equation}                       
where $\chi ({\bf R})$ is the free wave function of the hadron as a                       
function of the c.m. coordinate and $\psi ({\bf r})$ is the bound-state                       
solution of the $U(6/21)$ invariant Hamiltonian describing the                       
$q-\bar{q}$ mesons, $q-D$ baryons, $\bar{q}-\bar{D}$ antibaryons and                           
$D-\bar{D}$ exotic mesons, given by                       

\begin{equation}                      
i \partial_{t} \psi_{\alpha \beta} =                       
[\sqrt{(m_{\alpha} + \frac{1}{2} V_{s})^{2} + {\bf p}^{2}} +                      
\sqrt{(m_{\beta} + \frac{1}{2} V_{s})^{2} + {\bf p}^{2}} -\frac{4}{3}                      
\frac{\alpha_{s}}{r} + k \frac{{\bf s}_{\alpha} \cdot{\bf s}_{\beta}}                      
{m_{\alpha} m_{\beta}}] \psi_{\alpha \beta}                      
\end{equation}

Here ${\bf p}= -i {\bf \nabla}$ in the c.m. system and $m$ and $s$ denote                       
the masses and spins of the constituents, $\alpha_{s}$ the                       
strong-coupling constant, $V_{s} = br$ is the scalar potential with $r$                       
being the distance between the constituents in the bilocal object, and                        
$k= |\psi (0)|^{2}$.                      
                       
 The operator product expansion$^{\cite{wil}}$ will give a singular part depending only                        
on ${\bf r}$ and proportional to the propagator of the field binding the two                        
constituents.  There will be a finite number of singular coefficients $c_{n}                      
({\bf r})$ depending on the dimensionality of the constituent fields.                        
For example, for a                        
meson, the singular term is proportional to the progagator of the gluon field                        
binding the two constituents.  Once we subtract the singular part, the                        
remaining part $\tilde{O}({\bf R}, {\bf r})$ is analytic in r and thus                       
we can write

\begin{equation}                      
\tilde{O}({\bf R}, {\bf r})= O_{0}({\bf R}) + {\bf r} \cdot{\bf O}_{1}({\bf R})                      
+o(r^{2}).                      
\end{equation}                       
                       
   Now $O_{0}({\bf R})$ creates a hadron at its c.m. point ${\bf R}$                       
equivalent to a $\ell=0$, s-state of                        
the two constituents.  For a baryon this is a state associated with $q$ and                       
$D \sim qq$ at the same point ${\bf R}$, hence it is essentially a 3-quark                       
state when the three quarks are at a common location. The                        
${\bf O}_{1}({\bf R})$ can create three $\ell=1$ states with                        
opposite parity to the state created by $O_{0}({\bf R})$.  Hence, if                       
we denote the nonsingular parts of $\bar{q}(1)q(2)$ and $D(1)q(2)$ by                       
$[\bar{q}(1)q(2)]$ and $[D(1)q(2)]$, respectively, we have                       

\begin{equation}                      
[\bar{q}(1)q(2)]|0> = |M({\bf R})> + {\bf r} \cdot |{\bf M}^{'}({\bf R})> +                      
o(r^{2})                       
\end{equation}                       

\begin{equation}                      
[D(1)q(2)]|0> = |B({\bf R})> + {\bf r} \cdot |{\bf B}^{'}({\bf R})> +                      
o(r^{2})                       
\end{equation}
and similarly for the exotic meson states $D(1) \bar{D}(2)$.                      
                       
   Here $M$ belongs to the $(35+1)$-dimensional representation of $SU(6)$                        
corresponding to an $\ell=0$ bound state of the quark and the antiquark.                        
The $M^{'}({\bf R})$ is an orbital excitation $(\ell=1)$ of opposite                       
parity, which are in the $(35+1,3)$ representation of the group                       
$SU(6) \times O(3)$, $O(3)$ being associated with the relative                        
angular momentum of the constituents.  The $M^{'}$ states contain                       
mesons like $B$, ${\bf A}_{1}$, ${\bf A}_{2}$ and scalar particles.                        
On the whole, the $\ell=0$ and $\ell=1$ part $\bar{q}(1)q(2)$ contain                       
$4 \times(35+1) = 144$ meson states.                      
                       
   Switching to the baryon states, the requirement of antisymmetry in color                       
and symmetry in spin-flavor indices gives the $(56)^{+}$                       
representation for $B({\bf R})$. The $\ell=1$ multiplets have negative                       
parity and have mixed spin-flavor symmetry.                       
They belong to the representation $(70^{-},3)$ of $SU(6) \times O(3)$ and                       
are represented by the states $|{\bf B}^{'}({\bf R})>$ which are $210$ in                       
number.  On the whole, these $266$ states                       
account for all the observed low-lying baryon states obtained form                        
$56 + 3 \times 70 = 266$.  A similar analysis can be carried out for the                       
exotic meson states $D(1) \bar{D}(2)$, where the diquark and the                       
antidiquark can be bound in an $\ell=0$ or $\ell=1$ state with                       
opposite parities.

\begin{center}

{\bf  Color Algebra and Octonions}

\end{center}

The exact, unbroken color group
$SU(3)^c$ is the
backbone of the strong interaction. It is worthwhile to understand its role in the
diquark picture more clearly.

Two of the colored quarks in the baryon combine into an anti-triplet $ {\bf 3\times 3
=\bar{3}+(6)} $, $ {\bf 3\times \bar{3} = 1+(8)} $. The ${\bf (6)} $ partner of the diquark and the $ {\bf (8)} $ partner 
of the nucleon do not
exist. In hadron dynamics the only color combinations to consider
are $ {\bf 3\times 3 \rightarrow \bar{3}} $ and $ {\bf \bar{3}\times 3 \rightarrow 1} $. These relations imply 
the existence of split octonion units through
a representation of the Grassmann algebra $\{u_i,u_j\}=0$, $i= 1,2,3$. What is a bit strange is that 
operators $u_i$, unlike ordinary fermionic operators, are not associative. We also have 
$\frac{1}{2}[u_i,u_j]=\epsilon_{ijk}~u_k^{*}$. The Jacobi identity
does not hold since $[u_i,[u_j,u_k]]=- i e_7 \neq 0$, where $e_7$, anticommute with 
$u_i$ and $u_i^{*}$. 

~~~~  The behavior of various states under the color group are best                       
seen if we use split octonion units defined by$^{\cite{gug}}$                       

\begin{equation}                      
u_{0} = \frac{1}{2} (1 +i e_{7}) ,                      
~~~~~~~u_{0}^{*} = \frac{1}{2} (1 -i e_{7})                       
\end{equation}                        

\begin{equation}                      
u_{j} = \frac{1}{2} (e_{j} +i e_{j+3}) ,                      
~~~~~u_{j}^{*} = \frac{1}{2} (e_{j} -i e_{j+3}) , ~~~j=1,2,3                      
\end{equation}                       
                      
The automorphism group of the octonion algebra is the 14-parameter                       
exceptional group $G_{2}$.  The imaginary octonion units                       
$e_{\alpha} (\alpha  =1,...,7)$                       
fall into its 7-dimensional representation.                      
                       
   Under the $SU(3)^{c}$ subgroup of $G_{2}$ that leaves $e_{7}$                       
invariant, $u_{0}$ and $u_{0}^{*}$ are singlets, while $u_{j}$ and                       
$u_{j}^{*}$ correspond, respectively, to the                       
representations ${\bf 3}$ and $\bar{\bf 3}$.  The multiplication table can now be                       
written in a manifestly $SU(3)^{c}$ invariant manner (together with the                       
complex conjugate equations):

\begin{equation}                      
u_{0}^{2} = u_{0},~~~~~u_{0}u_{0}^{*} = 0                      
\end{equation}                       

\begin{equation}                      
u_{0} u_{j} = u_{j} u_{0}^{*} = u_{j},~~~~~                      
u_{0}^{*} u_{j} = u_{j} u_{0} = 0                          
\end{equation}                       

\begin{equation}                      
u_{i} u_{j}  = - u_{j} u_{i} = \epsilon_{ijk} u_{k}^{*}      \label{eq:oct}                      
\end{equation}                       

\begin{equation}                      
u_{i} u_{j}^{*} =  - \delta_{ij} u_{0}         \label{eq:octa}                      
\end{equation}                       
where $\epsilon_{ijk}$ is completely antisymmetric with  $\epsilon_{ijk} =1$                      
for  $ijk$  = $123$, $246$, $435$, $651$, $572$, $714$, $367$.                        
Here, one sees the                       
virtue of octonion multiplication.  If we consider the direct products

\begin{equation}                      
C:~~~~~{\bf 3} \otimes \bar{\bf 3} = {\bf 1} + {\bf 8}                       
\end{equation}

\begin{equation}                      
G:~~~~~{\bf 3} \otimes {\bf 3} = \bar{\bf 3} + {\bf 6}                      
\end{equation}                       
for $SU(3)^{c}$, then these equations show that octonion multiplication                       
gets rid of ${\bf 8}$ in ${\bf 3} \otimes \bar{\bf 3}$, while it gets rid                       
of ${\bf 6}$ in ${\bf 3} \otimes {\bf 3}$.  Combining  Eq.(\ref{eq:oct}) and                      
Wq.(\ref{eq:octa}) we find                       

\begin{equation}                      
(u_{i} u_{j}) u_{k} = - \epsilon_{ijk} u_{0}^{*}                       
\end{equation}                       
                       
   Thus the octonion product leaves only the color part in                       
${\bf 3} \otimes \bar{\bf 3}$ and ${\bf 3} \otimes {\bf 3} \otimes {\bf 3}$,                      
so that it is a natural algebra for colored quarks. 

For convenience we now produce the following multiplication table for the split octonion units:

\begin{center}
\begin{tabular}{|l|c|c|c|c|}\hline
 & $u_0$ & $u_{0}^{*}$ & $u_{k} $ & $u_{k}^{*}$  \\ \hline 
$u_0$  &  $ u_0 $  & $0$ & $u_k$ & $0$ \\ \hline

$u_{0}^{*}$  &  $ 0 $  & $u_{0}^{*}$ & $0$ & $u_{k}^{*}$ \\ \hline

$u_j$  &  $ 0 $  & $u_{j}$ & $\epsilon_{jki}u_{i}^{*}$ & $-\delta_{jk}u_{0}$ \\ \hline

$u_{j}^{*}$  &  $ u_{j}^{*} $  & $0$ & $-\delta_{jk}u_{0}^{*}$ & $\epsilon_{jki}u_{i}$ \\ \hline

\end{tabular}
\end{center}

It is worth noting that $u_i$ and $u_{j}^{*}$ behave like fermionic annihilation and creation operators:

\begin{equation}
\{u_i,u_j\}=\{u_{i}^{*},u_{j}^{*}\}=0,~~~\{u_i,u_{k}^{*}\}=-\delta_{ij}
\end{equation}                      

For more recent reviews on octonions and nonassociative algebras we refer to papers by Okubo$^{\cite{okubo}}$, and Baez$^{\cite{baez}}$. For generalized matrix representations of octonions we refer to Delbourgo, et al.$^{\cite{delbourgo}}$
                      
   The quarks, being in the triplet representation of the color                       
group $SU(3)^{c}$, are represented by the local fields                      
$q_{\alpha}^{i}(x)$, where $i = 1,2,3$ is the color index and $\alpha$                        
the combined spin-flavor index. Antiquarks at point $y$ are color                       
antitriplets $q_{\beta}^{i}(y)$.  Consider the two-body systems                       

\begin{equation}                      
C_{\alpha j}^{\beta i} = q_{\alpha}^{i} (x_{1}) \bar{q}_{\beta}^{j} (x_{2})                       
\label{eq:c}                      
\end{equation}                       

\begin{equation}                      
G_{\alpha \beta}^{i j} = q_{\alpha}^{i} (x_{1}) q_{\beta}^{j} (x_{2})                       
\label{eq:g}                      
\end{equation}                       
so that $C$ is either a color singlet or color octet, while $G$ is a                       
color antitriplet or a color sextet.  Now $C$ contains meson states                       
that are color singlets and hence observable.  The octet $q-\bar{q}$ state                       
is confined and not observed as a scattering state.  In the case of                       
two-body $G$ states, the antitriplets are diquarks which, inside a                       
hadron can be combined with another triplet quark to give                       
observable, color singlet, three-quark baryon states.  The color                       
sextet part of $G$ can only combine with a third quark to give                       
unobservable color octet and color decuplet three-quark states.                        
Hence the hadron dynamics is such that the ${\bf 8}$ part of $C$                       
and the ${\bf 6}$                       
part of $G$ are suppressed.  This can best be achieved by the use of the                       
above octonion algebra$^{ \cite{dom}}$.  The dynamical suppression of the                       
 octet and sextet states in Eq.(\ref{eq:c}) and Eq.(\ref{eq:g}) is , therefore,                       
automatically achieved.  The split octonion units can be contracted                       
with color indices of triplet or antitriplet fields.  For quarks                       
and antiquarks we can define the "transverse" octonions (calling $u_{0}$                       
and $u_{0}^{*}$ longitidunal units)

\begin{equation}                      
q_{\alpha} = u_{i} q_{\alpha}^{i} = {\bf u} \cdot{\bf q}_{\alpha} ,~~~~~                      
\bar{q}_{\beta} = u_{i}^{\dagger} \bar{q}_{\beta}^{j} = -{\bf u}^{*}                       
\cdot \bar{\bf q}_{\beta}        \label{eq:qal}                      
\end{equation}                       
                       
We find                       

\begin{equation}                      
q_{\alpha}(1) \bar{q}_{\beta}(2) = u_{0} {\bf q}_{\alpha}(1)                       
\cdot{\bf q}_{\beta}(2)                      
\end{equation}                       

\begin{equation}                      
\bar{q}_{\alpha}(1) q_{\beta}(2) = u_{0}^{*} \bar{\bf q}_{\alpha}(1)                       
\cdot{\bf q}_{\beta}(2)                      
\end{equation}

\begin{equation}                      
G_{\alpha \beta}(12) = q_{\alpha}(1) q_{\beta}(2) = {\bf u}^{*}                       
\cdot{\bf q}_{\alpha}(1) \times {\bf q}_{\beta}(2)                      
\end{equation}                       

\begin{equation}                      
G_{\beta \alpha}(21) = q_{\beta}(2) q_{\alpha}(1) = {\bf u}^{*}                       
\cdot{\bf q}_{\beta}(2) \times {\bf q}_{\alpha}(1)                      
\end{equation}                       
                       
Because of the anticomutativity of the quark fields, we have 
                      
\begin{equation}                      
G_{\alpha \beta}(12) = G_{\beta \alpha}(21) =                       
\frac{1}{2}  \{q_{\alpha}(1), q_{\beta}(2)\}                      
\end{equation}                       
                       
If the diquark forms a bound state represented by a field $D_{\alpha \beta}(x)$                      
at the center-of-mass location $x$

\begin{equation}                      
x = \frac{1}{2} (x_{1} +x_{2})                       
\end{equation}                       
when $x_{2}$ tends to $x_{1}$ we can replace the argument by $x$, and we obtain                       

\begin{equation}                      
D_{\alpha \beta}(x) = D_{\beta \alpha}(x)                      
\end{equation}                       
so that the local diquark field must be in a symmetric                       
representation of the spin-flavor group.  If the latter is taken to                       
be $SU(6)$, then $D_{\alpha \beta}(x)$ is in the 21-dimensional symmetric                       
representation, given by                        

\begin{equation}                      
({\bf 6} \otimes {\bf 6})_{s} = {\bf 21}                       
\end{equation}                       
                       
If we denote the antisymmetric $15$ representation by $\Delta_{\alpha \beta}$,                      
we see that the octonionic fields single out the $21$ diquark representation                       
at the expense of $\Delta_{\alpha \beta}$.  We note that without this                       
color algebra supersymmetry would give antisymmetric configurations as noted by                       
Salam and Strathdee$^{\cite{sal}}$ in their possible supersymmetric                       
generalization of hadronic supersymmetry.  Using the nonsingular                       
part of the operator product expansion we can write                       

\begin{equation}                      
\tilde{G}_{\alpha \beta}({\bf x}_{1}, {\bf x}_{2}) =                      
D_{\alpha \beta}(x) + {\bf r} \cdot {\bf \Delta}_{\alpha \beta}(x)                      
\label{eq:qam}                      
\end{equation}                       
The fields $\Delta_{\alpha \beta}$ have opposite parity to $D_{\alpha \beta}$;                      
${\bf r}$ is the relative                       
coordinate at time $t$ if we take $t$ = $t_{1}$ = $t_{2}$.  They play no role in                       
the excited baryon which becomes a bilocal system with the 21-                       
dimensional diquark as one of its constituents.                       
                      
   Now consider a three-quark system at time $t$.  The c.m. and                       
relative coordinates are

\begin{equation}                      
{\bf R} = \frac{1}{\sqrt{3}}({\bf r}_{1} + {\bf r}_{2} + {\bf r}_{3})                      
\end{equation}                       

\begin{equation}                      
{\bf  \rho} = \frac{1}{\sqrt{6}}(2 {\bf r}_{3} - {\bf r}_{1} - {\bf r}_{2})                     
\end{equation}                       

\begin{equation}                      
{\bf r} = \frac{1}{\sqrt{2}}({\bf r}_{1} - {\bf r}_{2})                      
\end{equation}                        
giving                       

\begin{equation}                      
{\bf r}_{1} = \frac{1}{\sqrt{3}} {\bf R} - \frac{1}{\sqrt{6}} {\bf \rho}                      
+ \frac{1}{\sqrt{2}} {\bf r}                      
\end{equation}                       

\begin{equation}                      
{\bf r}_{2} = \frac{1}{\sqrt{3}} {\bf R} - \frac{1}{\sqrt{6}} {\bf \rho}                      
- \frac{1}{\sqrt{2}} {\bf r}                      
\end{equation}                       

\begin{equation}                      
{\bf r}_{3} = \frac{1}{\sqrt{3}} {\bf R} + \frac{2}{\sqrt{6}} {\bf \rho}                      
\end{equation}                       
                       
The baryon state must be a color singlet, symmetric in the three                       
pairs ($\alpha$, $x_{1}$), ($\beta$, $x_{2}$), ($\gamma$, $x_{3}$).  We find                       

\begin{equation}                      
(q_{\alpha}(1) q_{\beta}(2)) q_{\gamma}(3) = -u_{0}^{*}                       
F_{\alpha \beta \gamma}(123)                       
\end{equation}                        

\begin{equation}                      
q_{\gamma}(3) (q_{\alpha}(1) q_{\beta}(2))  = -u_{0}                       
F_{\alpha \beta \gamma}(123)                      
\end{equation}                        
so that                       

\begin{equation}                      
- \frac{1}{2} \{\{q_{\alpha}(1), q_{\beta}(2)\}, q_{\gamma}(3)\} =                        
F_{\alpha \beta \gamma}(123)                       
\end{equation}                       
                       
The operator $F_{\alpha \beta \gamma}(123)$ is a color singlet and is                       
symmetrical in the three pairs of coordinates.  We have                       

\begin{equation}                      
F_{\alpha \beta \gamma}(123) = B_{\alpha \beta \gamma} ({\bf R}) +                      
{\bf \rho} \cdot {\bf B}'({\bf R}) + {\bf r} \cdot {\bf B}''({\bf R}) + C                      
\label{eq:fat}                      
\end{equation}                      
where $C$ is of order two and higher in ${\bf \rho}$ and ${\bf r}$.  Because                       
${\bf R}$ is symmetric in ${\bf r}_{1}$, ${\bf r}_{2}$ and ${\bf r}_{3}$,                       
the operator $B_{\alpha \beta \gamma}$    that creates a baryon                       
at ${\bf R}$ is totally symmetrical in its flavor-spin indices.  In the                       
$SU(6)$ scheme it belongs to the ($56$) representation.  In the bilocal                       
$q-D$ approximation we have ${\bf r}=0$ so that $F_{\alpha \beta \gamma}$ is                       
a function only of ${\bf R}$ and ${\bf \rho}$ which are both symmetrical in                       
${\bf r}_{1}$ and ${\bf r}_{2}$.  As before, ${\bf B}'$                        
belongs to the orbitally excited $70^{-}$ represenation of $SU(6)$.  The                       
totally antisymmetrical ($20$) is absent in the bilocal                       
approximation.  It would only appear in the trilocal treatment that                       
would involve the 15-dimensional diquarks.  Hence, if we use local                       
fields, any product of two octonionic quark fields gives a ($21$)                       
diquark                       

\begin{equation}                      
q_{\alpha}({\bf R}) q_{\beta}({\bf R}) = D_{\alpha \beta}({\bf R})                      
\end{equation}                       
and any nonassociative combination of three quarks, or a diquark                       
and a quark at the same point give a baryon in the $56^{+}$ representation:                       

\begin{equation}                      
(q_{\alpha}({\bf R}) q_{\beta}({\bf R})) q_{\gamma}({\bf R}) = - u_{0}^{*}                       
B_{\alpha \beta \gamma}({\bf R})                      
\end{equation}                       

\begin{equation}                      
q_{\alpha}({\bf R}) (q_{\beta}({\bf R}) q_{\gamma}({\bf R})) = - u_{0}                       
B_{\alpha \beta \gamma}({\bf R})                      
\end{equation}                       

\begin{equation}                      
q_{\gamma}({\bf R}) (q_{\alpha}({\bf R}) q_{\beta}({\bf R})) = - u_{0}                       
B_{\alpha \beta \gamma}({\bf R})                      
\end{equation}                       

\begin{equation}                      
(q_{\gamma}({\bf R}) q_{\alpha}({\bf R})) q_{\beta}({\bf R}) = - u_{0}^{*}                       
B_{\alpha \beta \gamma}({\bf R})                      
\end{equation}

The bilocal approximation gives the ($35+1$) mesons and the $70^{-}$                       
baryons with $\ell=1$ orbital excitation.                       
                      
\begin{center}
                      
{\bf  A Minimal Scheme}                      
   
\end{center}

 ~~~~  In order to go beyond the Miyazawa supersymmetry we now propose a nonsimple                      
super-Lie algebra that has parabolic structure with the odd part                      
being represented by a symmetric third rank tensor $F_{\alpha \beta \gamma}$                      
with $56$ components and the even part by the $36$  $U(6)$ generators                      
$L_{\sigma}^{\rho}$.  The bracket relations are                      

\begin{equation}                      
[L_{\sigma}^{\rho}, L_{\mu}^{\lambda}] = \delta_{\mu}^{\rho}                      
 L_{\sigma}^{\lambda}  - \delta_{\sigma}^{\lambda} L_{\mu}^{\rho}   \label{eq:l}                      
\end{equation}

\begin{equation}                      
\{F_{\alpha \beta \gamma}, F_{\rho \sigma \tau}\} = 0      \label{eq:la}                      
\end{equation}

\begin{equation}                      
[L_{\sigma}^{\rho}, F_{\alpha \beta \gamma}] = \delta_{\alpha}^{\rho}                      
F_{\beta \gamma \sigma} + \delta_{\beta}^{\rho} F_{\gamma \alpha \sigma} +                      
\delta_{\gamma}^{\rho} F_{\alpha \beta \sigma}            \label{eq:lb}                      
\end{equation}                       
                      
There is a conjugate superalgebra generated by $L_{\sigma}^{\rho}$  and                      
$F^{\alpha \beta \gamma}$.  Note that                      

\begin{equation}                      
N = \sum_{\rho} L_{\rho}^{\rho}                      
\end{equation}                       
acts like a number operator and traceless part of $L_{\sigma}^{\rho}$                      
coincides with                      
the elements of the $SU(6)$ Lie algebra.  Hence this minimal                      
superalgebra $S$ has $56 + 35 = 91$ generators.  With the addition of                      
color the algebra becomes $SU(3)^{c} \times S$.                      
                      
   We can immediately construct a matrix representation of this                      
algebra by means of fermionic operators $f_{\alpha}^{i}$ that transform like the                      
$(3, 6)$ representation of $SU(3)^{c} \times  SU(6)$ and the conjugate                       
operators $\bar{f}_{j}^{\beta}$ associated with the $(\bar{3}, \bar{6})$                       
representation.  We have                      

\begin{equation}                      
\{f_{\alpha}^{i}, f_{\beta}^{j}\} = 0, ~~~~~~~                      
\{\bar{f}_{i}^{\alpha}, \bar{f}_{j}^{\beta}\} = 0                      
\end{equation}                       

\begin{equation}                      
\{f_{\alpha}^{i}, \bar{f}_{j}^{\beta}\} = \delta_{j}^{i} \delta_{\alpha}^{\beta}                      
\end{equation}                      
                      
Since $f_{\alpha}^{i}$ are Grassmann numbers they can be represented by finite                      
matrices belonging to a Clifford algebra.  The conjugate operators                      
$\bar{f}_{\alpha}^{i}$                       
 are represented by the corresponding Hermitian conjugate matrices.                      
Out of this fermionic Heisenberg basis we construct the following matrices:                      

\begin{equation}                      
\Phi_{\alpha \beta \gamma} = \epsilon_{ijk} f_{\alpha}^{i} f_{\beta}^{j}                      
f_{\gamma}^{k}                       
\end{equation}                       

\begin{equation}                      
M_{\sigma}^{\rho} = \bar{f}_{i}^{\rho} f_{\sigma}^{i} -\frac{1}{6}                       
\delta_{\sigma}^{\rho} N                      
\end{equation}                       
where

\begin{equation}                      
N = \bar{f}_{j}^{\tau} f_{\tau}^{j}                      
\end{equation}                       
                      
Then $\Phi_{\alpha \beta \gamma}$  and $M_{\alpha}^{\rho}$ are the                       
representations of $F_{\alpha \beta \gamma}$  and $L_{\sigma}^{\rho}$                       
since they satisfy the algebra of Eqs.(\ref{eq:l}), (\ref{eq:la}) and                      
(\ref{eq:lb}). If $f_{\alpha}^{i}$ are identified with the                      
colored-quark annihilation operators, the odd elements of $\Phi$ of the                      
algebra correspond to baryons, while the even elements $M$ correspond                      
to mesons.  $N$ represents the number operator.                      
                         
At this point let us introduce a module that consists of the                      
representations $(\bar{3}, \bar{6})$ and $(\bar{3}, 21)$ of                       
$SU(3)^{c} \times SU(6)$.  They correspond to the operators associated                       
with the antiquarks and diquarks:                      

\begin{equation}                      
\bar{Q}_{j}^{\alpha} = \bar{f}_{j}^{\alpha},~~~~~~                      
D_{\alpha \beta k} = \epsilon_{ijk} f_{\alpha}^{i} f_{\beta}^{j}                      
\end{equation}                       
                      
   This multiplet is closed under the group $SU(3)^{c} \times S$.  Indeed we                      
have                      

\begin{equation}                      
\{\Phi_{\alpha \beta \gamma}, \bar{Q}_{k}^{\rho}\} = \delta_{\alpha}^{\rho}                      
D_{\beta \gamma k} + \delta_{\beta}^{\rho} D_{\gamma \alpha k}+                      
\delta_{\gamma}^{\rho} D_{\alpha \beta k}      \label{eq:pa}                      
\end{equation}                       

\begin{equation}                      
[\Phi_{\rho \sigma \tau} , D_{\alpha \beta k}] = 0      \label{eq:pb}                      
\end{equation}                       

\begin{equation}                      
[M_{\rho}^{\gamma}, \bar{Q}_{k}^{\sigma}] = \delta_{\rho}^{\sigma}                       
\bar{Q}_{k}^{\gamma}                      \label{eq:pc}                      
\end{equation}                       

\begin{equation}                      
[M_{\rho}^{\gamma}, D_{\alpha \beta k}] = \delta_{\alpha}^{\gamma}                      
D_{\beta \rho k} + \delta_{\beta}^{\gamma} D_{\alpha \rho k}                      
\label{eq:pd}                      
\end{equation}                       
                      
Eqs.(\ref{eq:pc}) and (\ref{eq:pd}) mean that $\bar{Q}$ and $D$ transform                      
respectively, like $ {\bf (\bar{6})} $ and $ {\bf (21)} $ under $SU(6)$.  Eqs.(\ref{eq:pa}) and (\ref{eq:pb})                      
tell us that the odd part of the supergroup maps $\bar{Q}$ into $D$ and $D$                      
into zero.  Letting the corresponding odd parameters be                       
$\eta^{\alpha \beta \gamma}$, we now                      
consider the superalgebra element                      

\begin{equation}                      
\Phi = \frac{1}{6} \eta^{\alpha \beta \gamma} \Phi_{\alpha \beta \gamma}                       
\end{equation}                       
                      
We find                      

\begin{equation}                      
\delta \bar{Q}_{k}^{\rho} = [\Phi , \bar{Q}_{k}^{\rho}] =                       
\frac{1}{2} \eta^{\rho \alpha \beta} D_{\alpha \beta k}                       
\end{equation}                       

\begin{equation}                      
\delta D_{\rho \sigma} = [\Phi, D_{\rho \sigma}] = 0                       
\end{equation}                       
                      
Neglecting mass and spin differences, the Hamiltonian we wrote down                      
earlier is invariant under the infinitesimal transformations                      
given above.                      
                      
   The baryon-meson system transforms like the adjoint                      
representation of the supergroup $S$.  We find                      

\begin{equation}                      
\delta B_{\alpha \beta \gamma} = [\Phi, B_{\alpha \beta \gamma}] = 0                        
\label{eq:ba}                      
\end{equation}                       

\begin{equation}                      
\delta M_{\sigma}^{\rho} = [\Phi , M_{\sigma}^{\rho}] =                       
- \frac{1}{2} \eta^{\rho \alpha \beta} B_{\alpha \beta \sigma}                       
\label{eq:bb}                      
\end{equation}                      
                      
Hence the odd part of $S$ maps mesons into baryons and baryons into                      
zero.  If the conjugate supergroup is used, baryons and diquarks                      
are replaced by antibaryons and antidiquarks, while the antiquarks                      
are replaced by quarks.                      
                      
   Note that in this minimal scheme the exotic mesons do not                      
appear.  Nevertheless the transformations Eq.(\ref{eq:ba}) and  Eq.(\ref{eq:bb}) are                      
sufficient for establishing the equality of the Regge slopes for                      
baryons and mesons.                      
                      
   Because of the form of the cubic function and the quadratic                      
functions of the generators of $S$ in terms of the fermionic                      
annihilation and creation operators, it is natural to introduce the                      
transverse octonion operators as in Eq.(\ref{eq:qal})                      

\begin{equation}                      
f_{\alpha} = u_{i} f_{\alpha}^{i},~~~~~~                      
\bar{f}^{\beta} = - u^{*j}  \bar{f}_{j}^{\beta}                        
\end{equation}                      
                      
Then we obtain in a general fashion                      

\begin{equation}                      
\Phi_{\alpha \beta \gamma} = \frac{1}{2} \{\{ f_{\alpha}, f_{\beta}\},                      
f_{\gamma}\}                       
\end{equation}                       
\begin{equation}                      
\{ \bar{f}^{\rho}, f_{\rho}\} = \frac{3}{2} \delta_{\sigma}^{\rho} +                      
i e_{7} M_{\sigma}^{\rho}                      
\end{equation}                       
                      
We also have                      

\begin{equation}                      
[f_{\alpha}, f_{\beta}] = 0, ~~~~~ [\bar{f}^{\alpha}, \bar{f}^{\beta}]=0                       
\label{eq:fa}                      
\end{equation}                       

\begin{equation}                      
[f_{\alpha}, \bar{f}^{\beta}] = M_{\alpha}^{\beta} + \frac{3}{2}                      
i e_{7} \delta_{\alpha}^{\beta}      \label{eq:fb}                            
\end{equation}                       
                      
Eqs.(\ref{eq:fa}) and (\ref{eq:fb}) together with the $SU(6)$ Lie algebra                      
of the form Eq.(\ref{eq:l}) satisfied by the operators $M_{\alpha}^{\beta}$                        
suggest that $M_{\alpha}^{\beta}$, $f_{\alpha}$, $\bar{f}^{\beta}$,                        
and  $i e_{7} = (u_{0} - u_{0}^{*})$ form an octonionic extension of $SU(6)$                       
that has $SU(6) \times SU(3)^{c}$ as a subgroup.  It is related to the                       
simple supergroup                      
$SU(6/3)$, but not equivalent to it because of the suppression of the                      
color octet and sextet states in the octonionic products                       
$f_{\alpha} \bar{f}^{\beta}$   and $f_{\alpha} f_{\beta}$.                      
                      
   The bilocal treatment outlined here also carries over                      
unchanged to the minimal scheme.  The $(\bar{q},D)$ system forms one                      
multiplet $R(1)$ at point $x_{1}$   under the supergroup $S$.  The conjugate                      
system $(q,\bar{D})$ forms a multiplet $R(2)$ at point $x_{2}$ under                       
the conjugate group $\bar{S}$.  Then the bilocal system                       
$R(1) \times R(2)$ is expanded as in Eq.(\ref{eq:qam}) and it transforms                       
under the group $S \times \bar{S}$.  Its local part with respect to c.m.                       
coordinate $x$ gives the meson-baryon and the meson-antibaryon multiplets                       
($35$, $56^{+}$) and ($35$, $56^{+}$), while the                      
coefficients of the relative coordinate ${\bf r}$ give the $\ell=1$ excitations                      
of mesons in the ($35+1$) representations and the $\ell=1$ excited baryons                      
in the $70^{-}$ representation of $SU(6)$.                      
                      
   In the symmetric approximation, the masses of various states are                      
proportional to the sum of the quark number plus the antiquark                      
number.  

\begin{center}

{\bf  Symmetries of the Three Quark System}

\end{center}

~~~~For the three quark system we can write the generalized form of the                 
two body Hamiltonian                   

\begin{eqnarray}                 
\tilde{H}_{123} & = & \tilde{V}_{12} + \tilde{V}_{23} + \tilde{V}_{31} +                  
\gamma_{4}^{(1)} \sqrt{[m_{1} + \frac{1}{2} S_{12}]^{2} +                  
\mbox{\boldmath $p$}_{12}^{2}}             \nonumber  \\    
                &   &                  
+ \gamma_{4}^{(2)} \sqrt{[m_{2} + \frac{1}{2} S_{12}]^{2} +                 
\mbox{\boldmath $p$}_{12}^{2}}        
  + \gamma_{4}^{(2)} \sqrt{[m_{2} + \frac{1}{2} S_{23}]^{2} +                  
\mbox{\boldmath $p$}_{23}^{2}}            \nonumber    \\      
                &   &   
+ \gamma_{4}^{(3)} \sqrt{[m_{3} + \frac{1}{2} S_{23}]^{2} +                 
\mbox{\boldmath $p$}_{23}^{2}}       
+ \gamma_{4}^{(3)} \sqrt{[m_{3} + \frac{1}{2} S_{31}]^{2} +                  
\mbox{\boldmath $p$}_{31}^{2}}              \nonumber    \\    
                &   &   
+ \gamma_{4}^{(1)} \sqrt{[m_{1} + \frac{1}{2} S_{31}]^{2} +                 
\mbox{\boldmath $p$}_{31}^{2}}                  
\end{eqnarray}                 
where                 
\begin{equation}                 
\tilde{V}_{ij} = V_{ij} + spin ~~dependent ~~terms                 
\end{equation}                 
with $i,j = 1, 2, 3$.                 
                 
To study the symmetry of this system we introduce fermion annihilation and                 
creation operators for colored quarks $f_{\alpha}^{i}$ with $i = 1, 2, 3$                  
the color index, and $\alpha = 1,\ldots,6$ the flavor index, so that 
                
\begin{equation}                 
\{f_{\alpha}^{i}, f_{\beta}^{j} \} = 0 , ~~~~                   
\{\bar{f}_{i}^{\alpha}, \bar{f}_{j}^{\beta}\} = 0 ,~~~~                 
\{f_{\alpha}^{i}, \bar{f}_{j}^{\beta} \} = \delta_{j}^{i}                  
\delta_{\alpha}^{\beta}                  
\end{equation}                 
                 
The diquark transforms like
                 
\begin{equation}                
D_{\alpha \beta k} = \epsilon_{ijk} f_{\alpha}^{i} f_{\beta}^{k}
\end{equation}                 
 It is in ($\bar{3}$) representation for color, and is in $21$ representation for $SU(6)$. The baryons transform like
                
\begin{equation}                
\Phi_{\alpha \beta \gamma} = \epsilon_{ijk} f_{\alpha}^{i}                
 f_{\beta}^{j} f_{\gamma}^{k}                
 \end{equation}                
It is a color singlet belonging to the $56$ representation of $SU(6)$. Color singlet 
mesons that are in the $35$ representation of $SU(6)$ transform like 
               
\begin{equation}                
M_{\sigma}^{\rho} = \bar{f}_{i}^{\rho} f_{\sigma}^{i} - \frac{1}{6}  \delta_{\sigma}^{\rho} N                
\end{equation}                
with $N$ being a color singlet and an $SU(6)$ singlet, and is given by  
              
\begin{equation}                
N = \bar{f}_{j}^{\tau} f_{\tau}^{j}                
\end{equation}
                 
 \begin{equation}           
[M_{\sigma}^{\rho}, M_{\mu}^{\lambda}] =            
\delta_{\mu}^{\rho} M_{\sigma}^{\lambda} -           
\delta_{\sigma}^{\lambda} M_{\mu}^{\rho}     \label{eq:su}           
\end{equation}
           
\begin{equation}           
\{ \Phi_{\alpha \beta \gamma}, \Phi_{\rho \sigma \tau} \} = 0           
                                                    \label{eq:sua}           
\end{equation}
           
\begin{equation}           
[M_{\sigma}^{\rho}, \Phi_{\rho \sigma \tau}] =            
\delta_{\alpha}^{\rho} \Phi_{\beta \gamma \sigma} +           
\delta_{\beta}^{\rho} \Phi_{\gamma \alpha \sigma} +           
\delta_{\gamma}^{\rho} \Phi_{\alpha \beta \sigma}     \label{eq:sub}           
\end{equation}           
           
Now $\bar{f}_{j}^{\rho}$ and $D_{\alpha \beta, k}$ form a multiplet           
that transforms under Eqs.(\ref{eq:su}), (\ref{eq:sua}), and (\ref{eq:sub})           
as follows: 
          
\begin{equation}           
\{ \Phi_{\alpha \beta \gamma}, \bar{f}_{k}^{\alpha} \} =           
\delta_{\alpha}^{\rho} D_{\beta \gamma, k} +           
\delta_{\beta}^{\rho} D_{\gamma \alpha, k} +           
\delta_{\gamma}^{\rho} D_{\alpha \beta, k}             
\end{equation}
           
\begin{equation}           
[\Phi_{\rho \sigma \tau}, D_{\alpha \beta, k}] = 0            
\end{equation} 
          
\begin{equation}           
[M_{\rho}^{\gamma}, \bar{f}_{k}^{\sigma}] =            
\delta_{\rho}^{\sigma} \bar{f}_{k}^{\gamma}            
\end{equation} 
          
\begin{equation}           
[M_{\rho}^{\gamma}, D_{\alpha \beta, k}] =           
\delta_{\alpha}^{\gamma} D_{\beta \rho, k} +           
\delta_{\beta}^{\gamma} D_{\alpha \rho, k}            
\end{equation}           
           
When the three quark system gets excited, it becomes a $q-D$ system,           
but retains its symmetry           
\begin{equation}           
B_{123} = \frac{1}{\sqrt{3}} (q_{1} D_{23} + q_{2} D_{31} + q_{3} D_{12})            
\end{equation}           
and remains in the $56$ representation. The $70$-representation has opposite           
parity (for example in $\mbox{\boldmath $r$}_{1} -          
\mbox{\boldmath $r$}_{23}$, etc.).            
           
When in the excited baryon quarks $2$ and $3$ form a diquark, and we can put 
          
\begin{equation}           
\mbox{\boldmath $p$}_{12} = \mbox{\boldmath $p$}_{13} =            
\mbox{\boldmath $p$} ~~~~~ \mbox{\boldmath $p$}_{23} = 0            
\end{equation}           
Then we obtain 
          
\begin{eqnarray}           
\tilde{H}_{123} \longrightarrow  & \tilde{H}_{q_{1} D_{23}}    
& \cong  2 \tilde{V}_{12} + \gamma_{4}^{(1)} \sqrt{[m_{1} +\frac{1}{2} S]^{2} +           
\mbox{\boldmath $p$}^{2}}  \nonumber   \\    
&        & + (\gamma_{4}^{(2)} +           
\gamma_{4}^{(3)}) \sqrt{[(m_{2} + m_{3} + \bar{S}_{23}) +\frac{1}{2} S]^{2} +           
\mbox{\boldmath $p$}^{2}}           
\end{eqnarray}           
and similarly for the other two configurations obtained by cyclic permutations.           
Note here that $ m_{D} = m_{2} + m_{3} + \bar{S}_{23}$ is the mass of the           
diquark.           
           
Then the superalgebra given by Eqs.(\ref{eq:su}), (\ref{eq:sua}) and           
(\ref{eq:sub}) which is symmetric  in ($123$) becomes, for a given $q-D$           
configuration the Miyazawa supersymmetry $1-(23)$.           
           
When $\tilde{H}_{123}$ is symmetrized it becomes invariant under the            
discrete group $S_{3}$. Hence its eigenstates are also eigenstates of            
$S_{3}$, i.e. correspond to representations of $S_{3}$. The splitting           
of $3$ quarks into $q$ and $D$ breaks the $S_{3}$ symmetry down to            
$Z_{2} \sim S_{2}$ which has two eigenstates (associated with $\pm 1$).           
Iachello$^{\cite{iach}}$ has shown that the $70$           
representation of $SU(6)$ associated with $\ell = 1$ is parity doubled           
while the ground state $56$ ($\ell = 0$) is not. This is supported by            
experiment.           

We can now define eight auxiliary octonions of quadratic norm $\frac{1}{2}$ 
constructed out of the seven imaginary units $f_{\alpha}$ ($\alpha = 1, \cdots 
7$).

\begin{equation}
f_{1} = \frac{1}{2} (e_{1}+e_{4}), ~~~~~~~
f_{2} = \frac{1}{2} (e_{2}+e_{5})  
\end{equation}

\begin{equation}
f_{3} = \frac{1}{2} (e_{3}+e_{6}), ~~~~~~~
f_{4} = \frac{1}{2} (e_{1}-e_{4}) 
\end{equation}

\begin{equation}
f_{5} = \frac{1}{2} (e_{2}-e_{5}), ~~~~~~~
f_{6} = \frac{1}{2} (e_{3}-e_{6}) 
\end{equation}

\begin{equation}
f_{7} = \frac{1}{2} (1-e_{7}), ~~~~
f_{8} = \frac{1}{2} (1+e_{7}) 
\end{equation}

These are intimately related to the eight split octonionic units and their 
conjugates we discussed above. If we keep one of the $f$'s fixed, say $f_{j}$, 
then the difference combination $(f_{i} f_{j} f_{k} - f_{k} f_{j} f_{i})$  is 
always either zero or equal to another $f$. We have worked out isomorphisms for 
such combination rules and they seem to play a fundamental role in dealing 
with symmetries of three quark systems we discussed, with further applications 
in multiquark systems. This will be discussed in detail in another 
publication$^{\cite{scbb}}$. 

Combinations of $f$'s also appear naturally within the 
root systems of groups associated with a magic square and play a profound role 
in superstring theories and their compactification. Generalization of split units and their implications to projective geometries and internal symmetries is discussed in our recent paper.$^{\cite{sultanc}}$

\begin{center}
                     
{\bf  A Colored Supersymmetry Scheme based on $SU(3)^{c} \times SU(6/1)$}

\end{center}

~~~~We could  go to a smaller supergroup having $SU(6)$ as a subgroup. With the addition of color, such a supergroup is $SU(3) \times SU(6/1)$. The fundamental representation of $SU(6/1)$ is $7$-dimensional which decomposes into a sextet and singlet under the spin-flavor group. There is also a $28$-dimensional representation of $SU(7)$. Under the $SU(6)$ subgroup it has the decomposition
                    
\begin{equation}                    
28 = 21 + 6 + 1                    
\end{equation}                    
                    
Hence, this supermultiplet can accommodate the bosonic antidiquark and fermionic quark in it, provided we are willing to add another scalar. Together with the color symmetry, we are led to consider the ($3$, $28$) representation of $SU(3) \times SU(6/1)$ which consists of an antidiquark, a quark and a color triplet scalar that we shall call a scalar quark. This boson is in some way analogous to the $s$ quarks. The whole multiplet can be represented by an octonionic $7 \times 7$ matrix $Z$ at point $x$.                    
                    
\begin{equation}                    
Z = {\bf u} \cdot                    
\left(                    
\begin{array}{cc}                    
\overline{\bf D}^{*} & {\bf q} \\                    
i {\bf q}^{T}  \sigma_{2} & {\bf S}                    
\end{array} \right)                    
\end{equation}                    
Here ${\bf D}^{*}$ is a $6 \times 6$ symmetric matrix representing the antidiquark, ${\bf q}$ is a $6 \times 1$ column matrix,             
${\bf q}^{T}$ is its transpose and $\sigma_{2}$  is the Pauli matrix that acts on the spin indices of the quark so that, if $q$ transforms with the $2 \times 2$ Lorentz matrix $L$,                     
${\bf q}^{T} i \sigma_{2}$ transforms with $L^{-1}$ acting from the right.                    
                    
Similarly we have                      

\begin{equation}                    
Z^{c} = {\bf u}^{*} \cdot                     
\left(                    
\begin{array}{cc}                    
{\bf D} & i \sigma_{2} {\bf q}^{*} \\                    
{\bf q}^{\dag} & {\bf S}^{*}                    
\end{array} \right)                    
\end{equation}                    
to represent the supermultiplet with a diquark and antidiquark. The mesons, exotic
 mesons and baryons are all in the bilocal field $Z(1) \otimes Z^{c}(2)$ which we expand 
with respect to the center of mass coordinates in order to represent color singlet hadrons 
by local fields. The color singlets $56^{+}$ and $70^{-}$ will then arise as in the earlier section.                    
                    
Now the ($\bar{D} q$) system belonged to the fundamental representation of the $SU(6/21)$  supergroup. But 
$Z$ belongs to the ($28$) representation of $SU(6/1)$ which is not its fundamental representation. Are there 
any fields that belong to the $7$-dimensional representation of $SU(6/1)$? It is possible to introduce such 
fictitious fields, such as a $6$-dimensional spinor $\xi$ and a scalar $a$ without necessarily assuming their existence as particles. We put                    
                    
\begin{equation}                    
\xi = {\bf u}^{*} \cdot {\bf \xi} , ~~~~~ a = {\bf u}^{*} \cdot {\bf a}                     
\end{equation}                    
so that both ${\bf \xi}$ and ${\bf a}$ are color antitriplets. Let 
                   
\begin{equation}                    
\lambda = \left(             
\begin{array}{c}             
\xi \\ a             
\end{array}                    
\right) ,~~~~~~                    
\lambda^{c} = \left(             
\begin{array}{c}             
\hat{\xi} \\ a^{*}             
\end{array}                    
\right)                    
\end{equation}                    
where 
                   
\begin{equation}                 
\hat{\xi} = {\bf u} \cdot (i \sigma_{2} {\bf \xi}^{*}) ,~~~~~                 
a^{*} = {\bf u} \cdot {\bf a}^{*}                  
\end{equation}                 
                 
Consider the $7 \times 7$ matrix 
                
\begin{equation}                 
W = \lambda \lambda^{c \dag} = \left(                  
\begin{array}{cc}                 
\xi \hat{\xi}^{\dag} & \xi a \\                 
a   \hat{\xi}^{\dag} & 0                 
\end{array}  \right)                  
\end{equation}                 
                 
$W$ belongs to the $28$-dimensional representation of $SU(6/1)$ and transforms like $Z$, provided the components of ${\bf \xi}$ are Grassmann numbers and ${\bf a}$ are even (bosonic) coordinates. The identification of $Z$ and $W$  give
                 
\begin{equation}                 
{\bf s} = {\bf a} \times {\bf a} = 0,~~{\bf q}_{\alpha} = {\bf \xi}_{\alpha} \times {\bf a},~~                 
D_{\alpha \beta}^{*} = {\bf \xi}_{\alpha} \times {\bf \xi}_{\beta}
\end{equation}               
                 
 A scalar part in $W$ can be generated by multiplying two different (7-dim) representations. The $56^{+}$ baryons form the color singlet part of the $84$-dimensional representation of $SU(6/1)$ while its colored part consists of quarks and diquarks.                 
                             
Now consider the octonionic valued quark field $q_{A}^{i}$, where    
$i = 1, 2, 3$ is the color index and $A$ stands for the pair    
$(\alpha, \mu)$ with $\alpha = 1, 2$ being the spin index and $\mu =    
1, 2, 3$ the flavor index. (If we have $N$ flavors, $A = 1, \ldots , N$).    
As before     
    
\begin{equation}    
q_{A} = u_{i} q_{A}^{i} = {\bf u} \cdot{\bf q}_{A}     
\end{equation}    
Similarly the diquark $D_{AB}$ which transforms like a color antitriplet    
is 
    
\begin{equation}    
D_{AB} = q_{A} q_{B} = q_{B} q_{A} = \epsilon_{ijk} u_{k}^{*}     
q_{A}^{i} q_{B}^{j} = {\bf u}^{*} \cdot{\bf D}_{AB}     
\end{equation}    
    
We note, once again, that because $q_{A}^{i}$ are anticommuting fermionic     
operators, $D_{AB}$ is symmetric in its two indices. The antiquark and     
antidiquark are represented by 
   
\begin{equation}    
\bar{q}_{A} = {\bf u}^{*} \cdot \bar{{\bf q}}_{A}    
\end{equation}   
and 
  
\begin{equation}    
\bar{D}_{AB} = {\bf u} \cdot \bar{{\bf D}}_{AB}    
\end{equation}    
respectively. If we have $3$ flavors, $q_{A}$ has $6$ components for each    
color while $D_{AB}$ has $21$ components.    
    
At this point let us study the system ($q_{A}$, $\bar{{\bf D}}_{BC}$)    
consisting of a quark and an antidiquark, both color triplets. There are     
two possibilities: we can regard the system as a multiplet belonging to     
the fundamental representation of a supergroup $U(6/21)$ for each color,    
or as a higher representation of a smaller supergroup. The latter possibility     
is more economical. To see what kind of supergroup we can have, we imagine    
that both quarks and diquarks are components of more elementary quantities:     
a triplet fermion $f_{A}^{i}$ and a boson $C^{i}$ which is a triplet    
with respect to the color group and a singlet with respect to $SU(2N)$    
($SU(6)$) for three flavors). The $f_{A}^{i}$ is taken to have baryon     
number $1/3$ while $C$ has baryon number $-2/3$. The system
     
\begin{equation}    
q_{A}^{i} = \epsilon_{ijk} \bar{f}_{A}^{j} \bar{C}^{k}    
\end{equation}    
will be a color triplet with baryon number $1/3$. It can therefore     
represent a quark. We can write  
  
\begin{equation}   
q_{A} = {\bf u} \cdot{\bf q}_{A} =   
( {\bf u}^{*} \cdot \bar{{\bf f}}_{A}) ( {\bf u}^{*} \cdot \bar{{\bf C}})   
\end{equation}   
With two anti-f fields we can form bosons that have same quantum numbers as   
antidiquarks:  
 
\begin{equation}   
\bar{D}_{AB} = {\bf u} \cdot \bar{{\bf D}}_{AB} =   
( {\bf u}^{*} \cdot \bar{{\bf f}}_{A}) ( {\bf u}^{*} \cdot \bar{{\bf f}}_{B})    
\end{equation}   
   
In this case the basic multiplet is $({\bf  f}_{A}, {\bf C})$ which belongs    
to the fundamental representation of $SU(6/1)$ for each color component. The   
complete algebra to consider is $SU(3) \times SU(6/1)$ and the basic multiplet    
corresponds to the representation $(3, 7)$ of this algebra. Let 
  
\begin{equation}   
F =    
\left(   
\begin{array}{c}   
f_{1} \\ f_{2} \\ \vdots \\ f_{6} \\ C   
\end{array} \right),   
~~~~~f_{A} = {\bf u} \cdot {\bf f}_{A}, ~~~~~ C = {\bf u} \cdot {\bf C}    
\end{equation}   
   
Also let 
  
\begin{equation}   
f^{1} =    
\left(   
\begin{array}{c}   
f_{1}^{1} \\ f_{2}^{1} \\ \vdots \\ f_{6}^{1}   
\end{array} \right),~~~~~   
f^{2} =    
\left(   
\begin{array}{c}   
f_{1}^{2} \\ f_{2}^{2} \\ \vdots \\ f_{6}^{2}   
\end{array} \right)   
\end{equation}   
   
Combining two such representations and writing $\bar{{\bf X}} = {\bf F} \times   
{\bf F}^{T}$ we have  
 
\begin{equation}   
\bar{X} = {\bf u}^{*} \cdot \bar{{\bf X}} =   
\left(   
\begin{array}{c}   
f^{1} \\ C^{1}   
\end{array} \right)   
\left(   
\begin{array}{cc}   
f^{2T} &   C^{2}   
\end{array} \right) -   
\left(   
\begin{array}{c}   
f^{2} \\ C^{2}   
\end{array} \right)   
\left(   
\begin{array}{cc}   
f^{1T} &   C^{1}   
\end{array} \right)   
\end{equation}   
   
Further identifying 
   
\begin{equation}   
{\bf u}^{*} \cdot {\bf D}_{11} = 2 f_{1}^{1} f_{1}^{2}, ~~~~~   
{\bf u}^{*} \cdot {\bf D}_{12} =  f_{1}^{1} f_{2}^{2} -   
f_{1}^{2} f_{2}^{1},~~  etc.,   
\end{equation}   
and   

\begin{equation}   
{\bf u}^{*} \cdot \bar{{\bf q}}_{1} = f_{1}^{1} C^{2} - f_{1}^{2} C^{1},~~~~   
{\bf u}^{*} \cdot \bar{{\bf q}}_{2} = f_{2}^{1} C^{2} - f_{2}^{2} C^{1},~~   
etc.,   
\end{equation}   
we see that $\bar{X}$ has the structure 
  
\begin{equation}   
\bar{X} = {\bf u}^{*} \cdot \bar{{\bf X}}    
 = {\bf u}^{*} \cdot   
\left(   
\begin{array}{cccc}   
{\bf D}_{11} & \ldots & {\bf D}_{16} & \bar{{\bf q}}_{1} \\   
{\bf D}_{12} & \ldots & {\bf D}_{26} & \bar{{\bf q}}_{2} \\   
{\bf D}_{13} & \ldots & {\bf D}_{36} & \bar{{\bf q}}_{3} \\   
{\bf D}_{14} & \ldots & {\bf D}_{46}  & \bar{{\bf q}}_{4} \\   
{\bf D}_{15} & \ldots & {\bf D}_{56}  & \bar{{\bf q}}_{5} \\   
{\bf D}_{16} & \ldots & {\bf D}_{66}  & \bar{{\bf q}}_{6} \\   
-\bar{{\bf q}}_{1} & \ldots & -\bar{{\bf q}}_{6} &  0            
\end{array} \right)   
\end{equation}   
or,
   
\begin{equation}   
( 3, 7) \times ( 3, 7) = (\bar{3} \times 27 )   
\end{equation}   
   
The $27$ dimensional representation decomposes into $21 + \bar{6}$ with   
respect to its $SU(6)$ subgroup.   
   
Consider now an antiquark-diquark system at point $x_{1} = x -   
\frac{1}{2} \xi$ and another quark-antidiquark system at point    
$x_{1} = x + \frac{1}{2} \xi$; Hence we take the direct product of    
$X(x_{1})$ and $X(x_{2})$. In other words 
  
\begin{equation}   
(D_{AB} (x_{1}), \bar{q}_{D} (x_{1})) \otimes   
(\bar{D}_{EF} (x_{2}), q_{C} (x_{2}))    
\end{equation}   
consisting of the pieces 
   
\begin{equation}   
H(x_{1}, x_{2}) = \left(   
\begin{array}{cc}   
\bar{q}_{D}(x_{1}) q_{C}(x_{2}) & D_{AB}(x_{1}) q_{C}(x_{2}) \\   
\bar{q}_{D}(x_{1}) \bar{D}_{EF}(x_{2}) & D_{AB}(x_{1}) \bar{D}_{EF}(x_{2})   
\end{array}  \right)   
\end{equation}   
   
The diagonal pieces are bilocal fields representing color singlet   
$1 + 35$ mesons and $1 + 35 + 405$ exotic mesons respectively with respect    
to the subgroup $SU(3)^{c} \times SU(6)$ of the algebra. The off diagonal   
pieces are color singlets that are completely symmetrical with respect to   
the indices $(ABC)$ and $(DEF)$. They correspond to baryons and antibaryons   
in the representations $56$ and $\bar{56}$ respectively of $SU(6)$.   
   
We can write $F_{ABC} = - \frac{1}{2} \{D_{AB}, q_{C}\}$ so that 
  
\begin{eqnarray}   
D_{AB} q_{C}&= & (u_{1}^{*} (q_{A}^{2} q_{B}^{3} + q_{B}^{2} q_{A}^{3}) +   
u_{2}^{*} (q_{A}^{3} q_{B}^{1} + q_{B}^{3} q_{A}^{1}) +   
u_{3}^{*} (q_{A}^{1} q_{B}^{2} + q_{B}^{1} q_{A}^{2})) \times  \nonumber  \\
            &  & 
(u_{1} q_{C}^{1} + u_{2} q_{C}^{2} + u_{3} q_{C}^{3})    
\end{eqnarray}   
becomes
   
\begin{equation}   
D_{AB} q_{C}=    
- u_{0}^{*} ( (q_{A}^{2} q_{B}^{3} + q_{B}^{2} q_{A}^{3}) q_{C}^{1} +   
(q_{A}^{3} q_{B}^{1} + q_{B}^{3} q_{A}^{1}) q_{C}^{2} +   
(q_{A}^{1} q_{B}^{2} + q_{B}^{1} q_{A}^{2}) q_{C}^{3})    
\end{equation}   
and similarly 
  
\begin{equation}   
q_{C} D_{AB} =    
- u_{0}  ((q_{A}^{2} q_{B}^{3} + q_{B}^{2} q_{A}^{3}) q_{C}^{1} +   
(q_{A}^{3} q_{B}^{1} + q_{B}^{3} q_{A}^{1}) q_{C}^{2} +   
(q_{A}^{1} q_{B}^{2} + q_{B}^{1} q_{A}^{2}) q_{C}^{3})   
\end{equation}   
Since $u_{0} + u_{0}^{*} = 1$, we have 
  
\begin{eqnarray}   
F_{ABC} = - \frac{1}{2} \{D_{AB}, q_{C}\}& = &  
(q_{A}^{1} q_{B}^{2} + q_{B}^{1} q_{A}^{2}) q_{C}^{3} +   
(q_{A}^{2} q_{B}^{3} + q_{B}^{2} q_{A}^{3}) q_{C}^{1} +  \nonumber   \\
                                         &   &
(q_{A}^{3} q_{B}^{1} + q_{B}^{3} q_{A}^{1}) q_{C}^{2}    
\end{eqnarray}   
which is completely symmetric with respect to indices $(ABC)$, corresponding    
to baryons.    
   
In the limit $x_{2} - x_{1} = \xi \longrightarrow 0$, $H$ can be represented   
by a local supermultiplet with dimension $ 2 \times 56 + 2(1 + 35) + 405 =   
589$ of the original algebra. This representation includes $56$ baryons,   
antibaryons, mesons and $q^{2} \bar{q}^{2}$ exotic mesons.   

\newpage 
 
\begin{center}
  
{\bf  Transformation Properties} 

\end{center}
  
 ~~Since $F = {\bf u} \cdot {\bf F}$ consists of three $7$-dimensional   
representations of $SU(6/1)$ we have  
 
\begin{equation}   
\delta F = Z~~F   
\end{equation}   
where $Z \in SU(6/1)$, 
  
\begin{equation}    
F =    
\left(   
\begin{array}{c}   
f \\ C   
\end{array} \right)   
=   
\left(   
\begin{array}{c}   
f_{1} \\ f_{2} \\ \vdots \\ f_{6} \\ C   
\end{array} \right)    
\end{equation}   
Hence $Z$ is a super antihermitian color singlet:  
 
\begin{equation}   
Z = \left(   
\begin{array}{cc}   
i H & \eta \\   
i \eta^{\dag} & i \omega   
\end{array}  \right)    
\end{equation}   
with $ H = H^{\dag}$, $ \omega = \omega^{*} $, $ Str Z = 0 $    
($Str$ =    
supertrace), $(\omega = tr H)$, and  
   
\begin{equation}   
\eta =    
\left(   
\begin{array}{c}   
\eta_{1} \\ \eta_{2} \\ \vdots \\ \eta_{6}   
\end{array} \right),   
~~~~~ \{ \eta_{\alpha}, \eta_{\beta} \} = 0    
\end{equation}   
$H$ is an antihermitian $6 \times 6$ matrix.    
   
Then by taking supertransposed quantities ($sT$ = supertransposed)
   
\begin{equation}   
\delta F^{sT} = \delta F^{T} = F^{T}~Z^{sT}   
\end{equation}   
we have 
 
\begin{equation}   
\left(   
\begin{array}{c}   
\delta {\bf f} \\  \delta {\bf C}   
\end{array} \right)   
=    
 \left(   
\begin{array}{cc}   
i H & \eta \\   
i \eta^{\dag} & i \omega   
\end{array}  \right)   
\left(   
\begin{array}{c}   
{\bf f} \\  {\bf C}   
\end{array} \right)    
\end{equation} 
  
\begin{equation}   
\left(   
\begin{array}{cc}   
\delta {\bf f}^{T} &  \delta {\bf C}   
\end{array} \right) =   
 \left(   
\begin{array}{cc}   
i H^{*} & -i \eta^{*} \\   
 \eta^{T} & i \omega   
\end{array}  \right)    
\end{equation}   
We also have 
  
\begin{equation}   
\bar{X} = F~~F^{T} = F~~F^{sT}   
\end{equation}   
so that  
 
\begin{equation}   
\delta \bar{X} = Z~~\bar{X} + \bar{X}~~Z^{sT}   
\end{equation}   
   
Now writing 
  
\begin{equation}   
\bar{X} =   
 \left(   
\begin{array}{cc}   
\bar{D} & q \\   
 - q^{T} & 0   
\end{array}  \right) ,    
~~~~~~(D = D^{T})    
\end{equation}   
under $U(6/21)$, $\delta X$ gives: 
  
\begin{equation}   
\delta \bar{D} = i (H \bar{D} + \bar{D} H^{T}) - (\eta q^{T} - q \eta^{T})    
\end{equation} 
  
\begin{equation}   
\delta q = i ( H + \omega ) q - \bar{D} \eta^{*}    
\end{equation}   
For supertransformation $SU(6/1)/U(6)$, the change in $\bar{D}$ and $q$ are   
   
\begin{equation}   
\delta \bar{D}=   q \eta^{T} - \eta q^{T}  ,~~~~~ \delta q =   
- \bar{D} \eta^{*}    
\end{equation}   
   
Since 
  
\begin{equation}   
\delta f = Z f    
\end{equation}   
or, in component form 
  
\begin{equation}   
\delta f_{A} = Z_{AB} f_{B}   
\end{equation}   
where $(A,B = 0, 1, \ldots, 6)$ with $f_{0} = C$, we have 
   
\begin{equation}   
\delta f^{sT*} = \delta f^{\dag} = f^{\dag}~Z^{*sT}    
\end{equation}   
   
If we define $g$ by 
  
\begin{equation}   
g =    
 \left(   
\begin{array}{cc}   
I  & 0 \\   
 0 &  i   
\end{array}  \right)     
\end{equation}   
then  
 
\begin{equation}   
\delta (f^{\dag}~g~f) =    
f^{\dag} (g Z + Z^{*sT} g) f    
\end{equation}   
It is easy to show $gZ + Z^{*sT} g = 0$ so that $\delta (f^{\dag}~g~f) = 0$.   
If we  look at $U(2/1)$ parts 
  
\begin{equation}   
\delta F =   
 \left(   
\begin{array}{cc}   
i H  & 0 \\   
 0 &  i \omega   
\end{array}  \right)   
\end{equation}   
giving  
 
\begin{equation}   
\delta f = i H f, ~~~~~ \delta C = i \omega C   
\end{equation}   
and   

\begin{equation}   
\delta F^{\dag} = F^{\dag}   
 \left(   
\begin{array}{cc}   
- i H  & 0 \\   
 0 & - i \omega   
\end{array}  \right)   
\end{equation}   
Since 
  
\begin{equation}   
\delta f^{\dag} = - i f^{\dag} H, ~~~~~ \delta C^{*} = - i \omega C^{*}    
\end{equation}   
we obtain  
 
\begin{equation}   
\delta (f^{\dag} f) = 0 
\end{equation}
and

\begin{equation}
 \delta (C^{*} C) = 0   
\end{equation}   
Similarly 
  
\begin{equation}   
\delta F =    
 \left(   
\begin{array}{cc}   
0  & \eta   \\   
 i \eta^{\dag} &  0   
\end{array}  \right)   
 F   
\end{equation}   
gives
   
\begin{equation}   
\delta f = \eta C, ~~~~~\delta C = i \eta^{\dag} f   
\end{equation}   
and using 
  
\begin{equation}   
\delta f^{\dag} = C^{*} \eta^{\dag}, ~~and~~   
\delta c^{*} = - i \eta^{T} f^{*} = i f^{\dag} \eta    
\end{equation}   
we arrive at 
   
\begin{equation}   
\delta (i f^{\dag} f - C^{\dag} C )= 0   
\end{equation}   
Defining $\bar{f} = i f^{\dag}$, we have  
 
\begin{equation}   
\delta (\bar{f} f - C^{\dag} C) = 0   
\end{equation}   
   
If we now define
    
\begin{equation}   
h_{AB \dot{A} \dot{B}} = (\bar{X})_{AB} (x) (X)_{\dot{A} \dot{B}} (x)   
\end{equation}   
where $A,B = 0, 1, \ldots, 6$ as before,  then the subset $h_{0b0\dot{b}}$   
antisymmetric in the first and last pairs of   
indices would describe $q \bar{q}$ mesons, $h_{ab\dot{a} \dot{b}}$   
symmetric in first and last pairs of indices  would describe    
 $q^{2} \bar{q}^{2}$ exotic mesons,  $h_{0b\dot{a} \dot{b}}$ antisymmetric   
in the first and symmetric in the last pair of indices would describe   
$q^{2} q$ baryons, and $h_{ab 0 \dot{b}}$ symmetric in the first and   
antisymmetric in the last pair of indices would describe   
$\bar{q}^{2} \bar{q}$ antibaryons.   
   
Aside from $h_{AB \dot{A} \dot{B}}$ describing baryons, antibaryons,   
mesons and exotics, this algebra can be extended to include preons $F_{A}$,   
antipreons $F_{\dot{A}}$, $X_{AB}$ describing $(\bar{q}^{2} q)$, and   
$X_{\dot{A} \dot{B}}$ describing $(q^{2} \bar{q})$. Gauge bosons and    
gauginos can be in the adjoint representation $V_{A \dot{B}}$. 
   
We note that, since 
  
\begin{equation}   
(Z_{AB})^{*} = (Z^{*})_{\dot{A} \dot{B}}   
\end{equation}   
we have 
  
\begin{equation}   
\delta F_{A} = Z_{AB}~F_{B} ~~and~~~   
\delta F_{\dot{A}} = Z_{AB}^{*}~F_{\dot{B}}     
\end{equation}   
 
Further applications of these extended algebraic structures and their implicaton for the construction of supersymmetric meson
baryon lagrangians will be dealt in another publication. 

\begin{center}

{\bf   Comments on Relativistic Formulation Through the Spin Realization of the 
Wess-Zumino Algebra}

\end{center}

It is possible to use a spin representation of the Wess-Zumino algebra 
to write first order relativistic equations for quarks and diquarks that are  
invariant under supersymmetry transformations. In this section we  
briefly deal 
with such Dirac-like supersymmetric equations and with a short discussion of  
experimental possibilities for the observation of the diquark structure 
and exotic $\bar{D}-D = (\bar{q}\bar{q})(qq)$ mesons. For a very nice  
discussion of the experimental situation we refer the reader to a  
recent papers by Anselmino, et.al.$^{ \cite{anse}}$, and by Klempt$^{\cite{klempt}}$. 
 
There is a spin realization of the Wess-Zumino super-Poincar\'{e} algebra 

\begin{equation} 
[p_{\mu}, p_{\nu}] = 0 ,~~~~~[D_{\alpha}, p_{\mu}] = 0 
\end{equation}
 
\begin{equation} 
[\bar{D}_{\dot{\beta}}, p_{\mu}] = 0, ~~~~~ 
[D^{\alpha}, \bar{D}^{\dot{\beta}}] = \sigma_{\mu}^{\alpha \dot{\beta}} 
p^{\mu} 
\end{equation} 
with $p_{\mu}$ transforming like a $4$-vector and $D^{\alpha}$, 
$\bar{D}^{\dot{\beta}}$ like the left and right handed spinors under the  
Lorentz group with generators $J_{\mu \nu}$. 
 
We also note that 

\begin{equation} 
[J_{\mu \nu}, p_{\lambda}] = \delta_{\mu \nu} p_{\lambda} - 
\delta_{\nu \lambda} p_{\mu} 
\end{equation} 
and 

\begin{equation} 
[J, J] = J  
\end{equation}  
 
The finite non unitary spin realization is in terms of $4 \times 4$ 
matrices for $J_{\mu \nu}$ and $p_{\nu}$ 

\begin{equation} 
J_{\mu \nu} = \frac{1}{2} \sigma_{\mu \nu} = 
\frac{1}{4i} [\gamma_{\mu}, \gamma_{\nu}]  
\end{equation} 

\begin{equation} 
J_{\mu \nu}^{L} = \frac{1-\gamma_{5}}{2} \frac{1}{2} \sigma_{\mu \nu} = 
\Sigma_{\mu \nu}^{L}    
\end{equation} 

\begin{equation} 
p_{\mu} = \Pi_{\mu}^{L} = \frac{1-\gamma_{5}}{2} \gamma_{\mu}   \label{eq:va} 
\end{equation} 
 
Introducing two Grassmann numbers $\theta_{\alpha}$ ($\alpha = 1, 2$) 
that transforms 
like the components of a left handed spinor and commute with the Dirac 
matrices $\gamma_{\mu}$, we have the representation 

\begin{equation} 
D_{\alpha} = \Delta_{\alpha} = \frac{\partial}{\partial \theta_{\alpha}} 
\end{equation}
 
\begin{equation} 
\bar{D}^{\dot{\beta}} = \bar{\Delta}^{\dot{\beta}} = \theta_{\alpha} 
\sigma_{\mu}^{\alpha \dot{\beta}} \Pi_{\mu}^{L}  
\end{equation} 
 
Such a representation of the super-Poincar\'{e} algebra acts on a Majorana  
chiral superfield 
\begin{equation} 
S(x, \theta) = \psi (x) + \theta_{\alpha} B^{\alpha} (x) + \frac{1}{2} 
\theta_{\alpha} \theta^{\alpha} \chi (x) . 
\end{equation} 
 
Here $\psi$ and $\chi$ are Majorana superfields associated with fermions and 
$B^{\alpha}$ has an unwritten Majorana index and a chiral spinor index 
$\alpha$, so that it represents a boson. 
 
Note that the sum of the two representations we wrote down is also a realization 
of the Wess-Zumino algebra. 
 
On the other hand we have the realization of $p_{\mu}$ in terms of the  
differential operator $-i \partial_{\mu} = -i \frac{\partial}{\partial  
x^{\mu}}$. In the Majorana representation, the operator $\gamma_{\mu} 
\partial_{\mu} = i \gamma_{\mu} p_{\mu}$ is real, and $\psi  = \psi ^{c} = 
\psi ^{*}$. Let us now define $\psi_{L}$ and $\psi_{R}$ by 

\begin{equation} 
\psi_{L} = \frac{1}{2} (1 + \gamma_{5}) \psi        
\end{equation}
and 

\begin{equation} 
\psi_{R} = \frac{1}{2} (1 - \gamma_{5}) \psi = \psi_{L}^{*}   
\end{equation} 
The free particle Dirac equation can now be written as 

\begin{equation} 
\Pi_{\mu}^{L} \partial_{\mu} \psi_{L} = m \psi_{L}^{*}    \label{eq:pi} 
\end{equation} 
or 

\begin{equation} 
\Pi_{\mu} p^{\mu} \psi_{L} = - i m \psi_{L}^{*} 
\end{equation} 
We can introduce left and right handed component fields 

\begin{equation} 
S_{L} = \frac{1}{2} (1 + \gamma_{5}) S  
\end{equation}  

\begin{equation} 
S_{R} = \frac{1}{2} (1 - \gamma_{5}) S = S_{L}^{*} 
\end{equation}
so that Eq.(\ref{eq:pi}) generalizes to the superfield equation 

\begin{equation} 
\Pi_{\mu}^{L} \partial_{\mu} S_{L} = m S_{L}^{*}    \label{eq:pia} 
\end{equation} 
or, 

\begin{equation} 
\Pi_{\mu} p^{\mu} S_{L} = - i m S_{L}^{*} 
\end{equation} 
Now consider the supersymmetry transformation 

\begin{equation} 
\delta S_{L} = (\xi^{\alpha} \Delta_{\alpha} + \bar{\xi}_{\dot{\beta}} 
\bar{\Delta}^{\dot{\beta}}) S_{L} = \Xi S_{L}    \label{eq:pib} 
\end{equation} 
This transformation commutes with the operator $\Pi_{\mu}^{L} \partial_{\mu}$ 
so that 

\begin{equation} 
\Pi_{\mu}^{L} \partial_{\mu} (S_{L} + \delta S_{L}) = m 
(S_{L} + \delta S_{L})^{*}. 
\end{equation} 
 
If $\psi_{L}$ is a left handed quark and $B^{\alpha}(x)$ an antidiquark 
with the same mass as the quark, Eq.(\ref{eq:pib}) provides a  
relativistic form of the quark antidiquark symmetry which is in fact broken 
by the quark-diquark mass difference. The scalar supersymmetric potential  
is introduced through $m \longrightarrow m + V_{s}$ as before and    
Eq.(\ref{eq:pia}) remains supersymmetric. By means of this formalism, it is 
possible to reformulate the treatments given in the earlier sections in  
first order relativistic form. 
 
To write equations in the first order form, we consider $V$ and $\Phi$ 
given in terms of the boson fields by 
\begin{equation} 
V = i \gamma_{\mu} V_{\mu} -\frac{1}{2}   \sigma_{\mu \nu} 
\end{equation} 
and 

\begin{equation} 
\Phi = i \gamma_{5} \phi + i \gamma_{5} \gamma_{\mu} \phi_{\mu} . 
\end{equation} 
In the Majorana representation 
 
\begin{equation} 
V^{*} = - V , ~~~~~and ~~~~~~ \Phi = \Phi^{*} 
\end{equation} 
 
We now define the left and right handed component fields by 
\begin{equation} 
V_{L} = \frac{1- \gamma_{5}}{2} V
\end{equation}
and

\begin{equation}
V_{R}= \frac{1+\gamma_{5}}{2} V 
\end{equation} 
so that

\begin{equation} 
V_{R} = - V_{L}^{*} 
\end{equation}  
with 

\begin{equation} 
V_{L} = i \frac{1- \gamma_{5}}{2} \gamma_{\mu} V_{\mu} - 
 \frac{1- \gamma_{5}}{2} \frac{1}{2} \sigma_{\mu \nu} V_{\mu \nu} 
\end{equation} 
and making use of  Eq.(\ref{eq:va}) we have
 
\begin{equation} 
V_{L} = i \Pi_{\mu}^{L} V_{\mu} - 
 \frac{1- \gamma_{5}}{2} \frac{1}{2} \sigma_{\mu \nu} V_{\mu \nu} 
\label{eq:vi} 
\end{equation} 
 
Noting that for any $a_{\mu}$ and $b_{\nu}$ we can write

\begin{equation} 
\Pi_{\mu} \Pi_{\nu}^{*} a_{\mu} b_{\nu} =  
\frac{1- \gamma_{5}}{2} \gamma_{\mu} \frac{1+ \gamma_{5}}{2} \gamma_{\nu} 
a_{\mu} b_{\nu} = \frac{1- \gamma_{5}}{2} \gamma_{\mu} \gamma_{\nu} 
a_{\mu} b_{\nu} 
\end{equation} 
and incorporating  the definition of $\sigma_{\mu \nu}$  

\begin{equation} 
\frac{1}{2} (\Pi_{\mu} \Pi_{\nu}^{*} - \Pi_{\nu} \Pi_{\mu}^{*}) = 
\frac{1- \gamma_{5}}{2} i \sigma_{\mu \nu} 
\end{equation} 
in Eq.(\ref{eq:vi}) leads to 

\begin{equation} 
V_{L} = i \Pi_{\mu}^{L} V_{\mu} + \frac{i}{2}  
(\Pi_{\mu} \Pi_{\nu}^{*} - \Pi_{\nu} \Pi_{\mu}^{*}) V_{\mu \nu}  
\label{eq:vl} 
\end{equation} 
Letting $\Sigma_{\mu \nu}^{L} = \Pi_{\mu} \Pi_{\nu}^{*} -  
\Pi_{\nu} \Pi_{\mu}^{*}$, Eq.(\ref{eq:vl}) reads 

\begin{equation} 
V_{L} = i \Pi_{\mu}^{L} V_{\mu} + \frac{i}{2} \Sigma_{\mu \nu}^{L} V_{\mu \nu}  
\end{equation} 

We now have a first order equation:
 
\begin{equation} 
\Pi_{\mu}^{L} \partial_{\mu} V_{R} = m V_{L}   \label{eq:vc} 
\end{equation} 
Similarly 

\begin{equation} 
\Pi_{\mu}^{R} \partial_{\mu} V_{L} = m V_{R} 
\end{equation} 
Therefore 

\begin{equation} 
\Pi_{\mu}^{L} \Pi_{\mu}^{R} \partial_{\mu} \partial_{\nu} V_{L} =  
 m \Pi_{\mu}^{L} \partial_{\mu} V_{R} = m^{2} V_{L} 
\end{equation} 
which after substitution of Eq.(\ref{eq:vc}) gives 

\begin{equation} 
\Box{V_{L}} = m^{2} V_{L} 
\end{equation} 

For the $\Phi$ part we can write 

\begin{equation} 
\Pi_{\mu}^{L} \partial_{\mu} \Phi_{R} = m \Phi_{L} 
\end{equation} 
with 

\begin{equation} 
\Phi_{L} = \frac{1- \gamma_{5}}{2} \Phi
\end{equation}
and

\begin{equation}
\Phi_{R} = \frac{1+ \gamma_{5}}{2} \Phi 
\end{equation} 

Clearly,

\begin{equation} 
\Phi_{R} = \Phi_{L}^{*}. 
\end{equation} 
and above procedure can now be repeated for the $\Phi$ fields.

\begin{center}
                                 
{\bf   Further Outlook}                           
 
\end{center}                   
~~~~ We have shown that the quark model with potentials derived                             
from QCD, including the quark-diquark model for excited hadrons, gives mass                             
formulae in very good agreement with experiment and goes a long way in                             
explaining the approximate symmetries and supersymmetries of the hadronic                            
 spectrum, including the symmetry-breaking mechanism.  For heavy quarks                             
the non-relativistic approximation can be used so that the potential models                             
for the spectra of charmonium$^{ \cite{ap}}$ and the                           
$b\bar{b}$ system$^{ \cite{eic}}$ are even simpler.  In this                             
approach gluons are eliminated leaving quarks interacting through potentials.                            
                          
     It is also possible to take an opposite approach by eliminating                             
quarks as well as gluons, leaving only an effective theory that involves                             
mesons (quark bound states) and baryons as collective excitations (solitons)                             
of the mesonic field.  This is the way pioneered by                           
Skyrme$^{ \cite{sky}}$ and revived in                             
recent years$^{ \cite{cho}}$.  The simplest model uses                           
an $SU(2) \times SU(2)$ nonlinear chiral                             
model involving pi-mesons as originally formulated by Skyrme in $1958$.  In                             
$1960$ Skyrme added a fourth order term to stabilize the soliton, so that it                             
could be interpreted as a nucleon.  The model was enriched by the Syracuse                             
school$^{ \cite{bal}}$ and by Witten and his collaborators$^{ \cite{adk}}$.  More recently vector mesons                            
 were added, improving the predictions$^{ \cite{ban}}$.  In the original model, mass                             
formulae were obtained only for a tower of $I = S$ excitations.  Recently                             
$SU(4)$ symmetry$^{ \cite{che}}$ and in principle $SU(6)$ symmetry were incorporated and                             
through fermionic quantization of the soliton the tower excitations were                             
chopped off, leading to a more realistic model$^{ \cite{cf}}$.                          
                            
     However, even in these improved versions the mass relations are not                             
nearly as good as in the potential quark model when the correct pion decay                             
constant is used$^{ \cite{rui}}$.  There is also an overall positive contribution to                             
hadronic masses that is not easily disposed of, although there are some                             
recent attemps to deal with the problem$^{ \cite{isl}}$.  In the absolute scale baryon                             
masses are about $20\%$ too high and vector mesons enter the theory on a very                             
different footing than pseudoscalar mesons, making it very difficult to                             
relate their masses.                            
                          
     A skyrme model that can compete with the potential quark model is                             
still hidden in the future. Partial solution to improved Skyrmion model was given in our recent paper,$^{\cite{catman}}$   suggesting a strategy towards building a full fledged quantitatively viable skyrmionic description of hadronic physics.                          
                            
     A better understanding of hadron masses through the QCD theory of                             
quarks and gluons has shifted the mass problem from the hadronic level                             
to the level of the elementary constituent quarks, leptons, weak bosons                             
and Higgs bosons of the standard model.  Since we now know that there are                             
three generations of fundamental fermions with light left handed neutrinos,                             
the development of a theory of masses for the finite system has become a                             
burning fundamental problem.  What is needed is the equivalent of a                             
Gell-Mann-Okubo$^{\cite{gel}}$ formula for the six quarks and for the leptons.  
A group theoretical treatment would have to involve a horizontal group associated with the 
three generations, namely a new $SU(3)$.  

     In order to approach the fundamental mass problem in imitation                             
of hadron masses we would have to introduce new primary particles  (preons) that would yield quarks and leptons as bound 
states.  But then it would be very difficult to understand why the fundamental   fermions of the standard model are finite in number.                            
 
\begin{center}

{\bf  Acknowledgments}

\end{center}

I thank Ramzi Khuri for discussions and a careful reading of the manuscript. Conversations with Vladimir Akulov, Anatoly Pashnev and Julius Wess have been particularly valuable.
                                

\newpage                               
                            
\begin{thebibliography}{99}
\bibitem{m1} H. Miyazawa, {\em Prog.\,Theor.\,Phys. {\bf 36} (1966) 1266;                                
Phys.\,Rev. {\bf 170} (1968) 1586.}                                
\bibitem{ram} P. Ramond, {\em Phys. Rev. {\bf{D3}} (1971)2415.}
\bibitem{nev} A. Neveu and J. Schwarz, {\em Nucl.Phys. {\bf {B31}} (1971) 86.}                            
\bibitem{gol} T.A. Golfand and E.P. Likhtman, {\em JETP Lett.{\bf{13}} (1971) 323.} 
\bibitem{vol} V.P. Akulov and D.V. Volkov, {\em Phys. Lett. {\bf{46b}}(1973) 109.}                            
\bibitem{wess} J. Wess and B. Zumino, {\em Nucl.Phys. {\bf {B70}} (1974) 39; 
Phys. Lett. {\bf{49b}} (1974) 52.}                            
\bibitem{geo} A. de Rujula, H. Georgi and S.L. Glashow, {\em Phys. Rev. {\bf {D12}} (1975) 147.}
\bibitem{l1} D.B. Lichtenberg, W. Namgung, E. Predazzi and J.G. Wills,                                
{\em Phys.\,Lett. {\bf 48} (1982) 1653}; W. Namgung and D.B. Lichtenberg,                                
{\em Lett.\,Nuovo Cimento {\bf 41} (1984) 597}; A. Martin, {\em Z.\, Phys. {\bf C32}                                 
(1986) 359}  and in "Symmetry in Nature" vol.II, p. 523 (Scuola Normale                                 
Superiore, Pisa, 1989).                                

\bibitem{anse} M. Anselmino, E. Predazzi, S. Ekelin, S. Fredriksson 
and D.B. Lichtenberg, {\em Rev. Mod. Phys. {\bf{65}} (1993) 1199}.                                                 

\bibitem{cg1} S. Catto and F. G\"{u}rsey, {\em Nuovo Cimento {\bf A 86} (1985) 201.}
\bibitem{egu} T. Eguchi, {\em Phys. Lett. B, {\bf 59} (1975) 457}.                                
\bibitem{jo} For strings as elongated bags see K. Johnson and                         
C.B. Thorn, {\em Phys. Rev. {\bf{D13}} (1976) 1934}.
\bibitem{cg2} S. Catto and F. G\"ursey, {\em Nuovo Cimento {\bf A99 } (1988) 685.}

\bibitem{d} K.N. Wilson, {\em Phys. Rev. {\bf D10} (1974) 2445.}
\bibitem{e} R.L. Jaffe, {\em Phys. Rev, {\bf D15} (1977) 281.} 
\bibitem{f} G. 't Hooft, {\em Nucl. Phys. {\bf B72} (1974) 461.} 
\bibitem{chodos} A. Chodos, R.L. Jaffe, K. Johnson and C.B. Thorn, {\em Phys. Rev. {\bf D10} (1974) 2599.} 

\bibitem{chodos2} T. De Grand, R.L. Jaffe, K. Johnson and J. Kiskis, {\em Phys. Rev. {\bf D12} (1975) 2060.} 

\bibitem{otto} S.W. Otto and J.D. Stark, {\em Phys. Rev. Lett. {\bf 52} (1984) 2328.} 
\bibitem{goddard} P. Goddard, J. Goldstone, C. Rebbi and C.B. Thorn, {\em Nucl. Phys. {\bf B56} (1973) 109.} 
\bibitem{chodos3} A. Chodos, et.al., {\em Phys. Rev. {\bf D9} (1974) 3471.} 
\bibitem{kac}For the classification of supergroups including U(m,n), see V.G. Kac, {\em Comm. Math. Phys. {\bf 53} (1977) 31.} 

                                
\bibitem{sbied} S. Catto, H.Y. Cheung and F. G\"ursey, {\em Mod. Phys. Lett. A,{\bf 6}, (1991) 3485.}
\bibitem{ma} A. Martin, Z.Phys.C32, 359(1986) and in "Symmetry in Nature",                             
vol.ii,p.523(Scuola Normale Superiore, Pisa, 1989).                            
                            
\bibitem{yam} H. Yamagishi, {\em Phys. Rev. {\bf D34} (1984) 269};  S. Takakura, T. Iwami and H. Kanada, {\em Prog.                             
Theor. Phys.{\bf 72} (1984) 379}; C.Habe et al. {\em Prog. Theor. Phy. {\bf 77} (1987) 917};                             
Z. Dziembowski and J. Franklin, {\em Phys. Rev. {\bf D42} (1990) 905}.                            

\bibitem{g3} C. Goebel, D. La Course and M.G. Olsson,                                 
{\em Phys.\,Rev. {\bf  D41} (1990) 2917}.
                                
\bibitem{i1} F. Iachello, {\em Nucl.\,Phys. {\bf A497} (1989) 23c};  F. Iachello,  N.C. Mukhopadhyay and L. Zhang, {\em 
Phys.\,Rev.\, {\bf B44} (1991) 898.}                                 


\bibitem{bidi} S. Catto, {\em Symmetries in Science VI.} (1993) 129. Ed. B. Gruber, Plenum Press.
\bibitem{gra} F. G\"{u}rsey and L.A. Radicati, {\em Phys. Rev. Lett. {\bf{13}} (1964) 299.} 
\bibitem{gre} O.W. Greenberg, {\em Phys. Rev. Lett. {\bf{13}} (1964) 598};                     
Y. Nambu in "Preludes in Theoretical Physics" eds. A. de Shalit,                     
H. Fesbach and L. van Hove, p.133 (Amsterdam, 1966);                     
H. Fritzsch, M. Gell-Mann and H. Leutwyler, {\em Phys. Lett. {\bf{47B}} (1973) 365}.                     
                     
\bibitem{tak} For ($qD$) system see S. Takakura, et al., {\em Prog. Theor. Phys. {\bf{77}} (1987) 917}.                     
                     
\bibitem{wil} K. Wilson, {\em Phys. Rev. {\bf 179} (1969) 1499}; and {\em Phys. Rev. {\bf{D3}} (1971) 1818.}                     
\bibitem{gug} M. G\"unayd{\i}n and F. G\"{u}rsey, {\em J. Math. Phys. {\bf 14} (1973) 1651.}
\bibitem{okubo} S. Okubo, {\em "Introduction to Octonion and other Nonassociative Algebras." Montroll Memorial 
Lecture Series in Mathematical Physics, 2. Cambridge Univ. Press (1995)}.
\bibitem{baez} J. Baez, {\em Bull. Am. Math. Soc. {\bf 39(2)} (2002) 145-205.}
\bibitem{delbourgo} J. Daboul and R. Delbourgo, {\em J. Math. Phys. {\bf 40} (1999) 4134}; arXiv:hep-th/9906065                                               
\bibitem{dom} Properties of this algebra is also discussed in G.                     
Domokos and S. K\"ovesi-Domokos, {\em J. Math. Phys. {\bf{19}} (1978) 1477.}                     
                     
\bibitem{sal} A. Salam and J. Strathdee, {\em Nucl. Phys. B{\bf 87} (1975) 85.}
\bibitem{iach} F. Iachello, {\em Lecture Notes in Physics {\bf{447}} (1995) 105. }
\bibitem{scbb} S. Catto, To be published.
\bibitem{sultanc} S. Catto, {\em Exceptional Projective Geometries and Internal Symmetries,"} arXiv:hep-th/0302079.                       
\bibitem{klempt} E. Klempt, {\em "Baryon Resonances and Strong QCD" arXiv:nucl-ex/0203002}.
\bibitem{ap} T.A. Appelquist and H.D. Politzer, {\em Phys.Rev.Lett. {\bf 34} (1975) 43.}                            
                            
\bibitem{eic} E. Eichten, et.al., {\em Phys.Rev. {\bf D21} (1980) 203.}                            
                            
\bibitem{sky} T.R.H. Skyrme, {\em Proc.Roy.Soc. {\bf A260} (1961) 127; Proc. Roy. Soc. {\bf A262} (1961) 237;                             
Nucl. Phys. {\bf 31} (1962) 566.}                            
                            
\bibitem{cho} See for a review A. Chodos, E. Hadjimichael and H.C. Tze                             
"Solitons in Nuclear and Elementary Particle Physics", (World                             
Scientific, Singapore, 1984).                            
                            
\bibitem{bal} A.P. Balachandran, V.P. Nair, S.G. Rajeev and A. Stern,                             
{\em Phys. Rev. Lett. {\bf 49} (1982) 1124; Phys. Rev. Lett. {\bf 50} (1983) 1620.}                            
                            
\bibitem{adk} G.S. Adkins, C.R. Nappi and E. Witten, {\em Nucl. Phys. {\bf 228} (1983) 522.}                            
                            
\bibitem{ban} For a review see M. Bando, T. Kugo and K. Yamawaki, {\em Phys. Rep. {\bf 164} (1988) 217.}                            
                            
\bibitem{che} H.Y. Cheung and F. G\"{u}rsey, {\em Phys. Lett. {\bf B219} (1989) 127;}  U.G. Meissner                             
and B.Pasquier, {\em Phys. Lett. {\bf 235} (1990) 153.}                            
                            
\bibitem{cf} H.Y. Cheung, Yale Ph.D. Thesis, 1991.                            
                            
\bibitem{rui} E.Ruiz Arriola, P. Alberto, K. Goeke and J.N. Urbano,                             
{\em Phys. Lett. {\bf 236} (1990) 381.}                            
                            
\bibitem{isl} M.M. Islam,  {\em Z. Phys. {\bf C53} (1992) 253.}                            

\bibitem{gel} M. Gell-Mann, {\em Phys. Rev. {\bf 125} (1962) 1062}; S. Okubo, 
{\em Prog. Theor. Phys. {\bf 27} (1962) 949.}                            
\bibitem{catman} S. Catto, {\em "Effective Hadronic Lagrangians Based on QCD: Potential Models and Skyrmions,"} arXiv:hep-th/0302101.






\end{thebibliography}
\end{document}